\documentclass[12pt]{article}
\usepackage{latexsym}
\usepackage{amsmath,amsfonts}
\usepackage{times}
\allowdisplaybreaks[4]

\catcode`\@=11

\newcount\hour
\newcount\minute
\newtoks\amorpm \hour=\time\divide\hour by 60\minute
=\time{\multiply\hour by 60 \global\advance\minute by-\hour}
\edef\standardtime{{\ifnum\hour<12 \global\amorpm={am}%
        \else\global\amorpm={pm}\advance\hour by-12 \fi
        \ifnum\hour=0 \hour=12 \fi
        \number\hour:\ifnum\minute<10
        0\fi\number\minute\the\amorpm}}
\edef\militarytime{\number\hour:\ifnum\minute<10 0\fi\number\minute}

\def\draftlabel#1{{\@bsphack\if@filesw {\let\thepage\relax
   \xdef\@gtempa{\write\@auxout{\string
      \newlabel{#1}{{\@currentlabel}{\thepage}}}}}\@gtempa
   \if@nobreak \ifvmode\nobreak\fi\fi\fi\@esphack}
        \gdef\@eqnlabel{#1}}
\def\@eqnlabel{}
\def\@vacuum{}
\def\marginnote#1{}
\def\draftmarginnote#1{\marginpar{\raggedright\scriptsize\tt#1}}
\def\draft{
        \pagestyle{plain}
        \overfullrule=2pt
        \oddsidemargin -.5truein
        \def\@oddhead{\sl \phantom{\today\quad\militarytime} \hfil
        \smash{\Large\sl DRAFT} \hfil \today\quad\militarytime}
        \let\@evenhead\@oddhead
        \let\label=\draftlabel
        \let\marginnote=\draftmarginnote
        \def\ps@empty{\let\@mkboth\@gobbletwo
        \def\@oddfoot{\hfil \smash{\Large\sl DRAFT} \hfil}
        \let\@evenfoot\@oddhead}
        \def\@eqnnum{(\theequation)\rlap{\kern\marginparsep\tt\@eqnlabel}%
        \global\let\@eqnlabel\@vacuum}  }

\newcommand{\rf}[1]{(\ref{#1})}
\renewcommand{\theequation}{\thesection.\arabic{equation}}
\renewcommand{\thefootnote}{\fnsymbol{footnote}}
\newcommand{\newsection}{   
\setcounter{equation}{0}\section}

\def\appendix#1{\addtocounter{section}{1}\setcounter{equation}{0}
\renewcommand{\thesection}{\Alph{section}}
\section*{Appendix \thesection\protect\indent \parbox[t]{11.15cm}{#1}}
\addcontentsline{toc}{section}{Appendix \thesection\ \ \ #1}}

\def\be{\begin{equation}}
\def\ee{\end{equation}}
\def\beq{\begin{eqnarray}}
\def\eeq{\end{eqnarray}}

\def\parline{\,\partial\kern -0.55em /\,\,}

\def\half{{\frac{1}{2}}}

\def\KK{{\cal K}}
\def\LL{{\cal L}}

\def\NN{{\cal N}}

\def\Fbf{{\bf F}}

\def\phibf{{\boldsymbol{\phi}}}

\def\noinbf#1{\noindent {\bf #1}}

\def\smCYM{{\scriptscriptstyle \rm CYM}}
\def\smHder{{\scriptscriptstyle \rm Hder}}

\def\(A)dS{{\rm (A)dS}}

\def\lin{{\rm lin}}
\def\tot{{\rm tot}}

\def\Tr{{\rm Tr}}
\def\ext{{\rm ext}}

\begin{document}


\begin{flushright}
FIAN-TD-2023-10  \hspace{2cm} \ \\
arXiv: 2309.12092 [hep-th],\  V2\\
\end{flushright}

\vspace{1cm}

\begin{center}

{\Large \bf Conformal Yang-Mills field in arbitrary dimensions }

\vspace{2.5cm}

R.R. Metsaev%
\footnote{ E-mail: metsaev@lpi.ru
}

\vspace{1cm}

{\it Department of Theoretical Physics, P.N. Lebedev Physical
Institute, \\ Leninsky prospect 53,  Moscow 119991, Russia }

\vspace{3.5cm}

{\bf Abstract}

\end{center}

Lagrangian of a classical conformal Yang-Mills field in the flat space of even dimension greater than or equal to six involves higher derivatives. We study Lagrangian formulation of the classical conformal Yang-Mills field by using ordinary-derivative (second-derivative) approach. In the framework of the ordinary-derivative approach, a field content, in addition to generic Yang-Mills field, consists of auxiliary vector fields and Stueckelberg scalar fields. For such field content, we obtain a gauge invariant Lagrangian with the conventional second-derivative kinetic terms and the corresponding gauge transformations. The Lagrangian is built in terms of non-abelian field strengths. Structure of a gauge algebra entering gauge symmetries of the conformal Yang-Mills field is described. FFF-vertex of the conformal Yang-Mills field which involves three derivatives is also obtained. For six, eight, and ten dimensions, eliminating the auxiliary vector fields and gauging away the Stueckelberg scalar fields, we obtain a higher-derivative Lagrangian of the conformal Yang-Mills field. For arbitrary dimensions, we demonstrate that all auxiliary fields can be integrated out at non-linear level leading just to a local higher-derivative action which is expressed only in terms of the generic Yang-Mills field.

\vspace{2.5cm}

\noindent Keywords: Conformal fields; Gauge invariant Lagrangian approach, Stueckelberg fields.

\noindent PACS: \ \ \ \ \ \ 11.25.Hf, 11.15.Kc

\newpage
\renewcommand{\thefootnote}{\arabic{footnote}}
\setcounter{footnote}{0}

\newsection{\large Introduction}

In view of various interesting features, conformal fields have attracted a lot of interest during long period of time (for review, see Ref.\cite{Fradkin:1985am}).
Conformal fields have been actively studied recently in the context of conformal supergravity theories. As an example we mention the investigations of $4d$, $\NN=4$ conformal supergravity by various methods in Refs.\cite{Buchbinder:2012uh} and $6d$ conformal supergravity in Refs.\cite{Linch:2012zh}.
Conformal fields have also been actively studied in the context of higher-spin field theories. For example, in the framework of AdS/CFT correspondence, for the free graviton field in Ref.\cite{Liu:1998ty} and for the free higher-spin field in Ref.\cite{Metsaev:2009ym}, it was shown that UV divergence of an action of AdS field computed on a solution of the Dirichlet problem coincides with the action of the  conformal field.%
\footnote{Recent interesting discussion of the AdS/CFT duality by using bi-local model may be found in Ref.\cite{deMelloKoch:2022sul}.
}
In the framework of induced action approach, the investigation of conformal higher-spin field action to fourth order in fields was carried out in Refs.\cite{Bekaert:2010ky,Beccaria:2016syk}. For more references on this topic, see Ref.\cite{Kuzenko:2022qeq}. For the recent study of conformal higher-spin gravity to all orders in fields and extensive list of references on this topic, see Ref.\cite{Basile:2022nou}, while, for the investigation of symmetry algebra of conformal higher-spin gravity, see Ref.\cite{Basile:2018eac}. Study of arbitrary spin conformal fields in (A)dS background and other curved backgrounds may be found in Refs.\cite{Joung:2012qy}-\cite{Manvelyan:2018qxd}.
Investigations of conformal higher-spin gravity in three dimensions may be found in  Refs.\cite{Grigoriev:2019xmp,Grigoriev:2020lzu,Ponomarev:2021xdq}. Application of conformal fields in the cosmological context may be found in Refs.\cite{Barvinsky:2023exr}.

In Refs.\cite{Metsaev:2007fq,Metsaev:2007rw}, we developed the ordinary-derivative Lorentz covariant and gauge invariant formulation of free arbitrary spin conformal fields in the $R^{d-1,1}$ space with even $d\geq 4$.%
\footnote{In Refs.\cite{Metsaev:2007fq,Metsaev:2007rw}, we dealt with short conformal fields. Extension of the ordinary-derivative approach to long and partial-short conformal free fields was obtained in Ref.\cite{Metsaev:2016oic}. Higher-derivative Lagrangian for all conformal fields of any symmetry was found in Ref.\cite{Vasiliev:2009ck} (for the related studies, see Refs.\cite{Shaynkman:2004vu,Dobrev:2012mea})
}
The most interesting feature of our approach is that kinetic terms of the conformal fields are realized as the conventional second-derivative kinetic terms. Namely, the kinetic terms for the  scalar, vector, and graviton conformal fields in $R^{d-1,1}$, $d\geq 4$, are realized as the respective Klein-Gordon, Maxwell, and Einstein-Hilbert kinetic terms. In this paper, we apply our ordinary-derivative approach for the study of interacting conformal Yang-Mills (YM) field in the $R^{d-1,1}$ space with even $d\geq 6$.

In Ref.\cite{Metsaev:2016rpa},\! using\! the\! ordinary-derivative\! light-cone\! gauge\! formulation of conformal fields, we have shown that, for the conformal YM field in $R^{d-1,1}$, $d\geq 6$, there exist only two cubic vertices: the first cubic vertex involves one derivative, while the second cubic vertex involves three derivatives. This implies that, in the framework of the ordinary-derivative Lorentz covariant and gauge invariant formulation of the conformal vector field in $R^{d-1,1}$, $d\geq 6$, there should be only two conformal invariants which, at third order in fields are associated with the two light-cone gauge cubic vertices in Ref.\cite{Metsaev:2016rpa}. {\it All other conformal invariants do not involve the second-order and third-order terms in the conformal YM field and they are beyond of the scope of the present paper}.

In order to formulate our aim in this paper more clearly we recall that, in the framework of standard higher-derivative approach, the most general Lagrangian of conformal YM field in $d$-dimensions can schematically be presented as
\be \label{manus-24102023-01}
\LL = I_2 + I_3 + \ldots + I_{d/2}\,, \qquad I_N : = \sum_{n=N}^{d/2} a_{N,n} D^{d-2n} \Fbf^n\,,
\ee
where $I_N$ is conformal invariant built in terms of field strength $\Fbf$ and covariant derivative $D$. Coefficients $a_{N,n}$ depend on coupling constants. The conformal invariant $I_N$ involves terms of the $N$th order and higher than the $N$th order in the YM field. In $6d$ there are the two well-known conformal invariants $I_2$ and $I_3$. To our knowledge, for $d>6$, the conformal invariants $I_N$, $N=2,3,\ldots,d/2$, have not been discussed in the earlier literature.

The aim of the present paper is to find two conformal invariants $I_2$ and $I_3$ for arbitrary $d\geq 6$ by using our ordinary-derivative approach. Using notation $F$ for a set of field strengths in our approach, we note the conformal invariant $I_2$ \rf{manus-24102023-01} takes the form $F^2$ and involves two derivatives. By analogy with the usual Poincar\'e invariant theory of YM field, we suggest refer this invariant to as Lagrangian of the conformal YM field. In our approach, the conformal invariant $I_3$ \rf{manus-24102023-01} takes the form $F^3$ and involves three derivatives. We refer the conformal invariant $I_3$ to as $F^3$-vertex,

Though our primary interest in this paper is the ordinary-derivative formulation of the conformal YM field we pay some attention to the standard higher-derivative approach to the conformal YM field. In the earlier literature, for $d\geq 6$, the conformal YM field was only discussed by using the higher-derivative formulation in $6d$.%
\footnote{ For $6d$ and $8d$ conformal gravities, the higher-derivative conformal invariants  were obtained in Refs.\cite{Bonora:1985cq,Boulanger:2004zf} (see also Refs.\cite{Boulanger:2004eh}). Application of ordinary-derivative approach for the study of $6d$ conformal gravity may be found in Ref.\cite{Metsaev:2010kp}, while application of unfolding conformal geometry approach for the study of $6d$, $8d$, and $10d$ conformal gravities may be found in Ref.\cite{Joung:2021bhf}. Various interesting aspects of $6d$ conformal gravity are discussed in Refs.\cite{Lu:2013hx}.
}
The most general commonly used higher-derivative Lagrangian of the conformal YM field in $6d$ is represented as $\LL=I_2+I_3$, where the conformal invariants $I_2$ and $I_3$ given in \rf{manus-24102023-01} involve four and three derivatives respectively.  Applying our approach to the conformal YM field in $6d$, we get these well known two conformal invariants. As we have already noted, for $d\geq 8$, conformal invariants of YM field have not been studied even in the framework of higher-derivative approach.
Therefore, for the conformal YM field in $8d$ and $10d$, we demonstrate, how two higher-derivative conformal invariants $I_2$ and $I_3$ can be obtained from our $F^2$ and $F^3$ conformal invariants.  For $d\geq 8$, all other higher-derivative conformal invariants $I_N$, $N>3$, do not involve the second-order and third-order terms in the conformal YM field and they are beyond of our study.

This paper is organized as follows.

In Sec.\ref{not-conv}, we collect our notation and conventions and outline a realization of conformal symmetries on fields and field strengths entering the ordinary-derivative approach.

In Sec.\ref{six}, as a warm-up, we apply our approach to the conformal YM field in $6d$. For such field, we present ordinary-derivative Lagrangian and $F^3$-vertex. Gauge transformations for fields entering the field content and for the corresponding fields strengths are discussed. Eliminating auxiliary fields and gauging away Stueckelberg fields, we show how our ordinary-derivative Lagrangian is related to the higher-derivative formulation of the conformal YM field in $6d$.

Secs.\ref{eight},\ref{ten} are devoted to the ordinary-derivative formulation of the conformal YM field in $8d$ and $10d$. Such dimensions are popular in various studies. Therefore we describe our results for $8d$ and $10d$ separately. We present the ordinary-derivative Lagrangian and the $F^3$-vertex of the conformal YM field. In $8d$, eliminating auxiliary fields and gauging away Stueckelberg fields, we show that our ordinary-derivative Lagrangian becomes the six-derivative conformal invariant (which we refer to as higher-derivative Lagrangian for the conformal YM field), while the $F^3$-vertex becomes the five-derivative conformal invariant. In $10d$, eliminating auxiliary fields and gauging away Stueckelberg fields, we show that our ordinary-derivative Lagrangian becomes the eight-derivative conformal invariant (which we refer to as higher-derivative Lagrangian for the conformal YM field), while the $F^3$-vertex becomes the seven-derivative conformal invariant.

In the main section of this article, Sec.\ref{arb}, we extend our results to the conformal YM field in the $R^{d-1,1}$ space with arbitrary even $d\geq 6$. For such field, we obtain the ordinary-derivative Lagrangian and the $F^3$-vertex. Gauge transformations and conformal algebra transformations for fields and the corresponding field strengths entering our formulation are also presented. Gauge algebra of the conformal YM field is also obtained. Also we demonstrate that all auxiliary fields can be integrated out at non-linear level leading just to a local higher-derivative action which is expressed only in terms of the field strength $F_2^{bc}$ and the covariant derivative $D^a$.

In Appendix A, we outline the derivation of non-abelian field strengths. In Appendices B, we outline the derivation of gauge transformations of fields and field strengths, while, in Appendix C, we discuss the derivation of the $F^3$-vertex.

\newsection{\large Conformal algebra. Notation and conventions } \label{not-conv}

\noinbf{Conformal algebra}. In $R^{d-1,1}$ space, the
conformal algebra $so(d,2)$ considered in a basis of the Lorentz algebra
$so(d-1,1)$ is decomposed into translation generators $P^a$, conformal boost
generators $K^a$, dilatation generator $D$ and generators of the Lorentz
algebra $so(d-1,1)$ denoted by $J^{ab}$. We use the following
commutation relations for the generators of the conformal algebra:
\beq
\label{manus-22082023-01} && [P^a,J^{bc}]=\eta^{ab}P^c -\eta^{ac}P^b\,,\hspace{1cm} [J^{ab},J^{ce}]=\eta^{bc}J^{ae}+3\hbox{ terms} \,,
\nonumber\\
&& [K^a,J^{bc}]=\eta^{ab}K^c - \eta^{ac}K^b\,, \hspace{1cm} [P^a,K^b]=\eta^{ab}D-J^{ab}\,,
\nonumber\\
&& [D,P^a]=-P^a\,, \hspace{3cm}  [D,K^a]=K^a\,.
\eeq
In \rf{manus-22082023-01} and below, the Lorentz algebra $so(d-1,1)$ vector indices take values $a,b,c,e,f =0,1,\ldots,d-1$. The flat space metric tensor $\eta^{ab}$ is assumed to be mostly positive. To simplify our
expressions we drop the metric tensor $\eta_{ab}$ in scalar products, i.e., we use   $X^aY^a \equiv \eta_{ab}X^a Y^b$.

\noinbf{Field content}. In Ref.\cite{Metsaev:2007fq}, we developed the ordinary-derivative gauge invariant Lagrangian formulation for free vector field having conformal dimension equal to one and propagating in the $R^{d-1,1}$ space.
In our approach, a field content entering the ordinary-derivative gauge invariant Lagrangian consists of the following set of vector and scalar fields of the Lorentz algebra $so(d-1,1)$:
\beq
\label{manus-22082023-02} && \phi_{2n+1}^a\,, \quad n=0,1,\ldots,\frac{d-4}{2}; \hspace{1.5cm} \phi_{2n}\,, \quad n=1,2,\ldots,\frac{d-4}{2}\,.\qquad
\\
\label{manus-22082023-03} && \Delta(\phi_{2n+1}^a) = 2n+1\,, \hspace{3cm} \Delta(\phi_{2n}) = 2n\,, \qquad
\eeq
where, in \rf{manus-22082023-03}, we show the conformal dimensions of the fields \rf{manus-22082023-02}, while the internal group indices of the fields are implicit. The vector field $\phi_1^a$ is a generic (primary) field which is identified with the field entering usual higher-derivative formulation of conformal vector field, while the remaining vector fields $\phi_{2n+1}^a$, $n=1,\ldots \frac{d-4}{2}$, turn out to be auxiliary fields required to get our ordinary-derivative Lagrangian formulation. All scalar fields $\phi_{2n}$ in \rf{manus-22082023-02} are Stueckelberg fields which are required to develop our ordinary-derivative gauge invariant formulation.

\noinbf{Realization of conformal algebra on fields and field strengths}. Let us use the shortcut $\Phi$ for fields \rf{manus-22082023-02}. In Ref.\cite{Metsaev:2007fq}, we found that, up to total derivatives,
our ordinary-derivative Lagrangian
is invariant with respect to the conformal algebra transformations given by
\be \label{manus-22082023-04}
\delta_G \Phi  = G \Phi \,,
\ee
where the realization of the conformal algebra generators on the field $\Phi$ takes the form
\beq
\label{manus-22082023-05} && P^a = \partial^a \,,
\nonumber\\
&& J^{ab} = x^a\partial^b -  x^b\partial^a + M^{ab}\,,
\nonumber\\
&& D = x^a\partial^a  + \Delta\,,
\nonumber\\
&& K^a = K_{\Delta,M}^a + R^a\,,
\nonumber\\
&&\qquad  K_{\Delta,M}^a \equiv -\frac{1}{2}x^bx^b\partial^a + x^a D + M^{ab}x^b\,, \qquad \partial^a=\eta^{ab}\partial/\partial x^b.
\eeq
In \rf{manus-22082023-05}, $\Delta$ is a conformal
dimension of the field $\Phi$, while $M^{ab}$ is a spin operator of the Lorentz algebra $so(d-1,1)$ which has the following  well-know realization
on the vector and scalar fields:
\be \label{manus-22082023-06}
M^{ab}\phi^c = \eta^{ac}\phi^b - \eta^{bc}\phi^a\,, \qquad
M^{ab}\phi = 0\,.
\ee
Expression for $K^a$ \rf{manus-22082023-05} implies that the conformal boost transformations can be presented as
\be \label{manus-22082023-07}
\delta_{K^a} \Phi = \delta_{K_{\Delta,M}^a} \Phi +
\delta_{R^a} \Phi \,,
\ee
where action of operator $K_{\Delta,M}^a$ \rf{manus-22082023-05} on the vector and scalar fields is given by
\beq
\label{manus-22082023-08} && \delta_{K_{\Delta,M}^a}\phi_{2n+1}^b = K_\Delta^a \phi_{2n+1}^b +
M^{abf} \phi_{2n+1}^f\,,
\nonumber\\
&& \delta_{K_{\Delta,M}^a} \phi_{2n} = K_\Delta^a \phi_{2n}\,,
\nonumber
\\
&& \hspace{2cm}  K_{\Delta}^a \equiv -\half x^bx^b
\partial^a + x^a (x^b\partial^b + \Delta) \,,
\hspace{1cm}  M^{abc} \equiv \eta^{ab}x ^c -\eta^{ac} x^b \,, \qquad
\eeq
while conformal dimensions of the fields are given in \rf{manus-22082023-03}. From \rf{manus-22082023-07} we see that in order to complete the description of the conformal boost transformations we should provide a realization of the operator $R^a$ on fields \rf{manus-22082023-02}. Schematically, the $R^a$-transformations of the fields  \rf{manus-22082023-02} can be presented as
\beq
\label{manus-22082023-09} &&  \phi_1^a \stackrel{R}{\longrightarrow} 0 \,,\hspace{1cm} \phi_{2n+1}^a \stackrel{R}{\longrightarrow} \phi_{2n}  \oplus
\partial \phi_{2n-1}^a\,,
\nonumber\\
&& \hspace{2.9cm} \phi_{2n} \stackrel{R}{\longrightarrow} \phi_{2n-1}^a \oplus
\partial \phi_{2n-2}\,,\qquad n=1,\ldots,\frac{d-4}{2}\,,
\eeq
where $\partial$ stands for derivative. From \rf{manus-22082023-09}, we see that the action of the operator $R^a$ on the generic (primary) field $\phi_1^a$ is trivial. For this reason, the operator $R^a$ does not appear in a standard higher-derivative formulation of conformal vector field. We see however that, in our approach, the action of the operator $R^a$ on the auxiliary fields and the Stueckelberg fields is non-trivial. Explicit realization  of the operator $R^a$ on the fields was found in Ref.\cite{Metsaev:2007fq} and will be presented below.

For fields \rf{manus-22082023-03}, we introduce the associated field strengths,
\beq
\label{manus-22082023-30}
&& F_{2n+2}^{bc}\,, \quad n=0,1,\ldots,\frac{d-4}{2}\,; \hspace{1.4cm} F_{2n+1}^b\,, \quad n=1,2,\ldots,\frac{d-4}{2}\,\,, \qquad
\\
&& \label{manus-22082023-31}  \Delta(F_{2n+2}^{bc}) = 2n+2\,, \hspace{3cm} \Delta(F_{2n+1}^b) = 2n+1\,, \qquad
\eeq
where, in \rf{manus-22082023-31}, we show the conformal dimensions of the field strengths \rf{manus-22082023-30}, while the internal group indices of the field strengths are implicit. Realization of the conformal algebra generators on the field strengths takes the same form as in \rf{manus-22082023-05}, where, in place of the relations \rf{manus-22082023-06}, we should use the following relations:
\be \label{manus-22082023-32}
M^{ab} F^{ce} = \eta^{ae} F^{cb} + 3 \hbox{ terms}\,, \qquad
M^{ab} F^c = \eta^{ac} F^b - \eta^{bc} F^a\,.
\ee
Action of conformal boost operator $K^a$ on the field strengths can be presented as
\be \label{manus-22082023-33}
\delta_{K^a} F^{bc} = K_{\Delta,M}^a F^{bc} +\delta_{R^a}F^{bc}\,, \qquad
\delta_{K^a} F^b = K_{\Delta,M}^a F^b + \delta_{R^a}F^b\,,
\ee
where the action of the operator $K_{\Delta,M}^a$ on the field strengths is given by
\beq
\label{manus-22082023-34} && \delta_{K_{\Delta,M}^a} F^{bc} = K_\Delta^a F^{bc} + M^{abf} F^{fc} + M^{acf} F^{bf}\,,
\nonumber\\
&& \delta_{K_{\Delta,M}^a} F^b = K_\Delta^a F^b + M^{abf} F^f\,,
\eeq
while the $K_\Delta^a$, $M^{abc}$ are defined in \rf{manus-22082023-08}. In this paper, we construct full non-abelian field strengths \rf{manus-22082023-30} in terms of the fields \rf{manus-22082023-02}.  Explicit expressions of the field strengths \rf{manus-22082023-30} in terms of the fields \rf{manus-22082023-02} as well as a realization of the operator $R^a$ on the field strengths we find in this paper will be given below.

We build our Lagrangian of the conformal YM field in terms of the field strengths \rf{manus-22082023-30}. Let us use the shortcut $F_n$ for the field strengths \rf{manus-22082023-30}. To simplify our expressions for the Lagrangian we use the sign $\,\,\sim\,\,$ to denote the trace operation over products of the field strengths,
\be \label{manus-22082023-50}
\LL \,\,\sim\,\, F_{n_1} \ldots F_{n_k} \qquad \Longleftrightarrow \qquad \LL = - \frac{\KK}{g_\smCYM^2}\Tr \big( F_{n_1} \ldots F_{n_k} \big)\,,
\ee
where $\Tr$ denotes the trace over a fundamental representation of an internal group,  $\KK$ a normalization factor, and $g_\smCYM$ is a dimensionless coupling constant of conformal YM field theory.%
\footnote{ For definiteness, the gauge group generators, $t^A$, $[t^A,t^B]=f^{ABC}t^C$, are considered in a fundamental representation of the internal symmetry group and normalized as $\Tr\, (t^A t^B) = - \KK^{-1}\delta^{AB}$, $\KK>0$. With the use of such conventions, our relation $\LL \,\,\sim\,\, - \frac{1}{4} F^{ab} F^{ab}$ leads to the commonly used relation $\LL = - \frac{1}{4g_\smCYM^2} (F^{ab})^A (F^{ab})^A$. In the context of higher-spin fields, the discussion of various internal symmetry groups may be found, e.g., in Refs.\cite{Konstein:1989ij,Metsaev:1991nb,Skvortsov:2020wtf}.
}
Explicit expressions for the Lagrangian will be given below.

\newsection{\large Conformal Yang-Mills field in six dimensions } \label{six}

{\bf Field content}. To discuss gauge invariant ordinary-derivative
formulation of the conformal YM field in $6d$ flat space we use two  vector fields and
one scalar field:
\beq
\label{manus-16082023-01} && \phi_1^a \qquad\quad \phi_3^a
\nonumber\\
&&
\\[-25pt]
&& \hspace{0.9cm} \phi_2
\nonumber
\eeq
The fields $\phi_1^a$, $\phi_3^a$ are vector fields of the Lorentz algebra $so(5,1)$, while the field $\phi_2$ is a scalar field of the $so(5,1)$.  Conformal dimensions of the fields in \rf{manus-16082023-01} are given by
\be \label{manus-16082023-02}
\Delta(\phi_1^a) = 1\,,\qquad \Delta(\phi_3^a) = 3\,,\qquad \Delta(\phi_2) = 2\,.
\ee
Note that the field $\phi_1^a$ has the same conformal dimension as a conformal YM field entering the higher-derivative approach. This is to say that in our approach the field $\phi_1^a$ is identified with the conformal YM field, while the field $\phi_3^a$ turns out to be auxiliary field. As we show below, the scalar field $\phi_2$ is realized as a Stueckelberg field in our approach.

\noinbf{Gauge invariant Lagrangian}. We use the sign $\,\,\sim\,\,$ defined in \rf{manus-22082023-50}. The ordinary-derivative Lagrangian of the conformal YM field we found can then be presented in terms of field strengths as
\be \label{manus-16082023-03}
\LL_\smCYM \,\,\sim\,\, - \half F_2^{ab} F_4^{ab}  -  \half F_3^a F_3^a \,,
\ee
where the field strengths are given by
\beq
\label{manus-16082023-04} && F_2^{ab} = \partial^a \phi_1^b - \partial^b \phi_1^a + [\phi_1^a,\phi_1^b]\,,
\nonumber\\
&& F_4^{ab} = D^a \phi_3^b - D^b \phi_3^a \,,
\nonumber\\
&& F_3^a =  \phi_3^a + D^a \phi_2 \,,
\eeq
while the covariant derivative $D^a$ of the fields $\phi_3^a$ and $\phi_2$  is defined as
\be \label{manus-16082023-05}
D^a \phi_3^b :=  \partial^a \phi_3^b +  [\phi_1^a, \phi_3^b] \,,\hspace{2cm} D^a \phi_2 := \partial^a \phi_2 + [\phi_1^a,\phi_2] \,.
\ee
From \rf{manus-16082023-04}, we see that $F_2^{ab}$ is a field strength for the conformal YM field $\phi_1^a$, while, from \rf{manus-16082023-05}, we learn that the  field $\phi_1^a$ is realized as the YM connection of the covariant derivative $D^a$.

\noinbf{Gauge transformations}. To discuss gauge symmetries of Lagrangian \rf{manus-16082023-03} we introduce the following set of gauge parameters:
\be \label{manus-16082023-06}
\xi_0\,, \qquad \xi_2\,. \qquad
\ee
Gauge parameters \rf{manus-16082023-06} are scalar fields of the Lorentz algebra $so(5,1)$. Gauge transformations of the fields take the following form:
\beq
\label{manus-16082023-07} && \delta \phi_1^a = D^a \xi_0 \,,
\nonumber\\
&& \delta \phi_3^a = D^a \xi_2 +  [\phi_3^a,\xi_0]\,,
\nonumber\\
&& \delta \phi_2 = -\xi_2 + [\phi_2,\xi_0]\,,
\eeq
where action of the covariant derivative $D^a$ on gauge parameters is given by
\be \label{manus-16082023-08}
D^a\xi_0 := \partial^a \xi_0 + [\phi_1^a,\xi_0]\,, \qquad  D^a\xi_2 := \partial^a \xi_2 +   [\phi_1^a,\xi_2]\,.
\ee
Gauge transformations \rf{manus-16082023-07} and definitions \rf{manus-16082023-04} lead to the following gauge transformations of the field strengths:
\beq
\label{manus-16082023-09} && \delta F_2^{ab} =   [F_2^{ab},\xi_0]\,,
\nonumber\\
&& \delta F_4^{ab} =  [F_4^{ab},\xi_0] + [F_2^{ab},\xi_2]\,,
\nonumber\\
&& \delta F_3^a =  [F_3^a,\xi_0]\,.
\eeq

\noinbf{Gauge algebra}. Let $\xi_0$, $\xi_2$ and $\eta_0$, $\eta_2$ be two sets of gauge transformation parameters. Using gauge transformations \rf{manus-16082023-07}, we find the following commutators of two gauge transformations:
\beq
\label{manus-16082023-10} && [\delta_{\eta_0},\delta_{\xi_0}] =   \delta_{ [\eta_0,\xi_0] }\,,
\hspace{1cm}  [\delta_{\eta_0},\delta_{\xi_2}] =   \delta_{ [\eta_0,\xi_2] }\,,
\nonumber\\
&& [\delta_{\eta_2},\delta_{\xi_2}] =  0\,.
\eeq
Using \rf{manus-16082023-10} it is easy to verify that the Jacobi identities for three gauge transformations are satisfied.

\noinbf{Conformal transformations of fields and field strengths}. Transformations of  fields \rf{manus-16082023-01} under action of the operator $R^a$ take the following form:
\beq
\label{manus-16082023-30} && \delta_{R^a}\phi_1^b  = 0  \,,
\nonumber\\
&& \delta_{R^a}\phi_3^b  =  - 2\eta^{ab} \phi_2 -2\partial^a \phi_1^b\,,
\nonumber\\
&& \delta_{R^a} \phi_2  =  2\phi_1^a \,.
\eeq
Using the relations given in \rf{manus-16082023-04} and \rf{manus-16082023-30}, we find the following $R^a$-transformations of the field strengths:
\beq
\label{manus-16082023-31} && \delta_{R^a} F_2^{bc}  = 0\,,
\nonumber\\
&& \delta_{R^a} F_4^{bc}  = 2\eta^{ab} F_3^c - 2\eta^{ac} F_3^b  - 2 \partial^a  F_2^{bc} \,,
\nonumber\\
&& \delta_{R^a} F_3^b  =   - 2 F_2^{ab}\,.
\eeq
The following remarks are in order.

\noinbf{i}) Lagrangian  \rf{manus-16082023-03} does not involve higher than second-order terms in the derivatives. Two-derivative contributions to Lagrangian \rf{manus-16082023-03} take the following form (up-to normalization factors):
\be
\phi_3^a (\eta^{ab} \Box  - \partial^a\partial^b)\phi_1^b\,, \hspace{1cm} \phi_2 \Box \phi_2\,.
\ee
We see that, for the vector fields, the two-derivative contribution to Lagrangian \rf{manus-16082023-03} takes the form of the standard Maxwell kinetic term, while, for the scalar field, the two-derivative contribution to Lagrangian \rf{manus-16082023-03} takes the form of the standard Klein-Gordon kinetic term.

\noinbf{ii}) From gauge transformations \rf{manus-16082023-07}, we see that the vector fields are realized as gauge fields.

\noinbf{iii}) Also from gauge transformations \rf{manus-16082023-07}, we see that
the scalar field $\phi_2$ transforms as Stueckelberg (Goldstone) fields under the gauge transformations. In other words, the scalar field $\phi_2$ is realized as Stueckelberg field in our approach.

\noinbf{iv}) Lagrangian \rf{manus-16082023-03} is the only ordinary-derivative Lorentz invariant functional that can be built in terms of the field strengths. We verify that  Lagrangian \rf{manus-16082023-03} is invariant under gauge transformations \rf{manus-16082023-09} and conformal algebra transformations described in \rf{manus-22082023-05}, \rf{manus-22082023-33}, and \rf{manus-16082023-31}.

\noinbf{v}) We note certain similarities between the ordinary-derivative Lagrangian for conformal fields and gauge invariant Lagrangian for massive fields. For various gauge invariant formulations of massive arbitrary spin fields and list of references, see, e.g., Refs.\cite{Zinoviev:2008ze,Khabarov:2021xts}.

\noinbf{$F^3$-vertex of conformal YM field}. In the framework of our approach, $F^3$-vertex of the conformal YM field in $6d$ coincides with the well-known $F^3$-vertex which we denote as $\LL_{F^3}$,
\be \label{manus-16082023-45}
\LL_{F^3} \,\,\sim\,\,  g F_2^{ab} F_2^{bc} F_2^{ca}\,,
\ee
where the field strength $F_2^{ab}$ is defined in \rf{manus-16082023-04}, while the sign $\,\,\sim\,\,$ is defined in  \rf{manus-22082023-50}.  The $g$ is a dimensionless coupling constant. Needless to say, vertex $\LL_{F^3}$ \rf{manus-16082023-45} does not involve higher than third-order terms in the derivatives. We now demonstrate the interrelation between our formulation and the standard higher-derivative formulation.

\noinbf{ Higher-derivative Lagrangian of conformal YM field}. Gauging away the Stueckelberg scalar field $\phi_2$ and ignoring total derivatives, we represent  ordinary-derivative Lagrangian \rf{manus-16082023-03} as
\be \label{manus-16082023-70}
\LL_\smCYM \,\,\sim\,\, \phi_3^a D^bF_2^{ba} - \half \phi_3^a \phi_3^a \,, \qquad D^b F_2^{bc} := \partial^b F_2^{bc} + [\phi_1^b,F_2^{bc}]\,.
\ee
Using equations of motion for the auxiliary vector field $\phi_3^a$ obtained from Lagrangian \rf{manus-16082023-70}, we find the solution for the auxiliary vector field $\phi_3^a$, which we denote as $\phibf_3^a$,
\be \label{manus-16082023-71}
\phibf_3^a = D^b F_2^{ba}\,.
\ee
Plugging solution $\phibf_3^a$ \rf{manus-16082023-71} into ordinary-derivative Lagrangian \rf{manus-16082023-70}, we get the well-known higher-derivative Lagrangian of the conformal YM field, which we denote as $\LL_\smCYM^\smHder$,
\be \label{manus-16082023-72}
\LL_\smCYM^\smHder \,\,\sim\,\, \half D^bF_2^{ba} D^cF_2^{ca}\,,
\ee
where $F_2^{ab}$ is defined in \rf{manus-16082023-04}. Lagrangian $\LL_\smCYM^\smHder$   \rf{manus-16082023-72} does not involve higher than fourth-order terms in the derivatives. Conformal invariance of $\LL_\smCYM^\smHder$ \rf{manus-16082023-72} is verified by using $\delta_{R^a}\phibf_3^b = -2F^{ab}$.

\noinbf{ General theory of conformal YM field}. The most general Lagrangian of the conformal YM field in $6d$ is defined to be
\be \label{manus-16082023-87}
\LL_\tot = \LL_\smCYM + \LL_{F^3}\,,
\ee
where $\LL_\smCYM$, $\LL_{F^3}$ are given in \rf{manus-16082023-03}, \rf{manus-16082023-45}. As is well known, for YM field in $6d$, there are only two conformal invariants. In the framework of ordinary-derivative approach it is the Lagrangian $\LL_\tot$ that describes those two conformal invariants.

For the illustration purposes, we consider higher-derivative cousin of Lagrangian \rf{manus-16082023-87}. Namely, gauging away the Stueckelberg scalar field $\phi_2$ and using equations of motion for the auxiliary vector field $\phi_3^a$ obtained from Lagrangian \rf{manus-16082023-87}, we find the solution for the $\phi_3^a$ which we denote as $\phibf_3^a$. The solution $\phibf_3^a$ takes the form given in \rf{manus-16082023-71}. Plugging the $\phibf_3^a$ into the Lagrangian $\LL_\tot$ \rf{manus-16082023-87}, we get
\be  \label{manus-16082023-88}
\LL_\tot^\smHder = \LL_\smCYM^\smHder + \LL_{F^3}^\smHder\,, \qquad \LL_{F^3}^\smHder \,\,\sim\,\, g F_2^{ab} F_2^{bc} F_2^{ca}\,,
\ee
where $\LL_\smCYM^\smHder$ is given in \rf{manus-16082023-72}, while $F_2^{ab}$ is  given in \rf{manus-16082023-04}. It is the higher-derivative Lagrangian $\LL_\tot^\smHder$ \rf{manus-16082023-88} that is well known and commonly used in the literature when discussing the general conformal invariant theory of the YM field in $6d$.

\newsection{\large Conformal Yang-Mills field in eight dimensions } \label{eight}

{\bf Field content}. To discuss gauge invariant ordinary-derivative
formulation of the conformal YM field in $8d$ flat space we use three  vector fields and
two scalar fields:
\beq
\label{manus-17082023-01}  && \phi_1^a \qquad\quad \phi_3^a \qquad\quad \phi_5^a
\nonumber\\
&&
\\[-25pt]
&& \hspace{0.8cm} \phi_2 \qquad \quad \phi_4
\nonumber
\eeq
The fields $\phi_1^a$, $\phi_3^a$, and $\phi_5^a$ are vector fields of the Lorentz algebra $so(7,1)$, while the fields $\phi_2$ and $\phi_4$ are scalar fields of the $so(7,1)$.  Conformal dimensions of the fields in \rf{manus-17082023-01} are given by
\be \label{manus-17082023-02}
\Delta(\phi_{2n+1}^a) = 2n+1\,, \hspace{0.5cm} n=0,1,2\,; \hspace{2cm} \Delta(\phi_{2n}) = 2n\,, \hspace{0.7cm} n=1,2\,.\qquad
\ee
The vector field $\phi_1^a$ has the same conformal dimension as the conformal YM field  entering the higher-derivative approach. Therefore the vector field $\phi_1^a$ is identified with the conformal YM field, while the vector fields $\phi_3^a$, $\phi_5^a$ turn out to be auxiliary fields. Below we show that the scalar fields $\phi_2$, $\phi_4$ are realized as Stueckelberg fields in our approach.

\noinbf{Gauge invariant Lagrangian}.We use the sign $\,\,\sim\,\,$ defined in \rf{manus-22082023-50}. The ordinary-derivative Lagrangian of the conformal YM field we found can then be presented in terms of field strengths as
\be \label{manus-17082023-03}
\LL_\smCYM  \,\,\sim\,\, -\half F_2^{ab} F_6^{ab} - \frac{1}{4}F_4^{ab} F_4^{ab} -   F_3^a F_5^a\,,
\ee
where the field strengths are given by
\beq
\label{manus-17082023-04} && F_2^{ab} = \partial^a \phi_1^b - \partial^b \phi_1^a +  [\phi_1^a,\phi_1^b]\,,
\nonumber\\
&& F_4^{ab} = D^a \phi_3^b - D^b \phi_3^a\,,
\nonumber\\
&& F_6^{ab}  = D^a \phi_5^b - D^b \phi_5^a + [\phi_3^a,\phi_3^b]\,,
\nonumber\\[5pt]
&& F_3^a = \phi_3^a + D^a\phi_2\,,
\nonumber\\
&& F_5^a = \phi_5^a + D^a\phi_4  + \half [\phi_3^a,\phi_2]\,,
\eeq
while action of the covariant derivative $D^a$ on the auxiliary fields $\phi_3^a$, $\phi_5^a$ and the Stueckelberg fields $\phi_2$, $\phi_4$  is defined as
\be \label{manus-17082023-05}
D^a\phi_{2n+1}^b := \partial^a \phi_{2n+1}^b + [\phi_1^a,\phi_{2n+1}^b]\,, \qquad D^a\phi_{2n} := \partial^a \phi_{2n} + [\phi_1^a,\phi_{2n}]\,, \hspace{0.5cm} n = 1,2\,.
\ee
From \rf{manus-17082023-04}, we see that $F_2^{ab}$ is a field strength of the conformal YM field $\phi_1^a$, while, from \rf{manus-17082023-05}, we see that the field $\phi_1^a$ is realized as the YM connection entering the covariant derivative $D^a$.

\noinbf{Gauge transformations}. To discuss gauge symmetries of Lagrangian \rf{manus-17082023-03} we introduce the following set of gauge parameters:
\be \label{manus-17082023-06}
\xi_0\,, \qquad \xi_2\,, \qquad \xi_4\,. \qquad
\ee
Gauge parameters \rf{manus-17082023-06} are scalar fields of the Lorentz algebra $so(7,1)$. Gauge transformations take the following form:
\beq
\label{manus-17082023-07} && \delta_\xi \phi_1^a  = D^a \xi_0\,,
\nonumber\\
&& \delta_\xi \phi_3^a  = D^a \xi_2 + [\phi_3^a,\xi_0] \,,
\nonumber\\
&& \delta_\xi \phi_5^a  = D^a \xi_4 + [\phi_5^a,\xi_0] + [\phi_3^a,\xi_2] \,,
\nonumber\\
&& \delta_\xi \phi_2 = - \xi_2 + [\phi_2,\xi_0] \,,
\nonumber\\
&& \delta_\xi \phi_4 = - \xi_4 + [\phi_4,\xi_0] + \half   [\phi_2,\xi_2] \,,
\eeq
where action of the covariant derivative $D^a$ on gauge parameters \rf{manus-17082023-06} is given by
\be
D^a\xi_{2n} := \partial^a \xi_{2n} + [\phi_1^a,\xi_{2n}]\,.
\ee
Using relations \rf{manus-17082023-04}, \rf{manus-17082023-07}, we find that gauge transformations of the field strengths take the form
\beq
\label{manus-17082023-08} && \delta_\xi F_2^{ab} =  [F_2^{ab},\xi_0]\,,
\nonumber\\
&& \delta_\xi F_4^{ab} = [F_4^{ab},\xi_0] + [F_2^{ab},\xi_2]\,,
\nonumber\\
&& \delta_\xi F_6^{ab} =  [F_6^{ab},\xi_0] + [F_4^{ab},\xi_2] + [F_2^{ab},\xi_4] \,,
\nonumber\\
&& \delta_\xi F_3^a =  [F_3^a,\xi_0]\,,
\nonumber\\
&& \delta_\xi F_5^a = [F_5^a,\xi_0] + \half [F_3^a,\xi_2] \,.
\eeq

\noinbf{Gauge algebra}. Let $\xi_n$ and $\eta_n$, $n=0,2,4$, be two sets of gauge parameters. Using gauge transformations \rf{manus-17082023-07}, we find the following commutators of two gauge transformations:
\beq
\label{manus-17082023-09} && [\delta_{\eta_0},\delta_{\xi_0}] =   \delta_{ [\eta_0,\xi_0] }\,,
\hspace{1cm} [\delta_{\eta_0},\delta_{\xi_2}] =   \delta_{ [\eta_0,\xi_2] }\,,
\hspace{1cm}  [\delta_{\eta_0},\delta_{\xi_4}] =   \delta_{ [\xi_0,\xi_4] }\,,
\nonumber\\
&& [\delta_{\eta_2},\delta_{\xi_2}] =   \delta_{ [\eta_2,\xi_2] }\,,
\hspace{1cm}  [\delta_{\eta_2},\delta_{\xi_4}] = 0\,,
\nonumber\\
&& [\delta_{\eta_4},\delta_{\xi_4}] = 0\,.
\eeq
Using \rf{manus-17082023-09}, we verify that the Jacobi identities for three gauge transformations are satisfied.

\noinbf{$R^a$-transformations of fields and field strengths}. Transformations of fields \rf{manus-17082023-01} under action of the operator $R^a$ take the following form:
\beq
\label{manus-17082023-30} &&  \delta_{R^a} \phi_1^b  = 0\,,
\nonumber\\
&&  \delta_{R^a} \phi_3^b  =  - 2\eta^{ab} \phi_2 - 4\partial^a \phi_1^b\,,
\nonumber\\
&&  \delta_{R^a} \phi_5^b  =  - 4\eta^{ab} \phi_4 - 4 \partial^a \phi_3^b\,,
\nonumber\\
&&  \delta_{R^a} \phi_2  =  4 \phi_1^a \,,
\nonumber\\
&& \delta_{R^a} \phi_4  =   2 \phi_3^a - 2\partial^a \phi_2\,.
\eeq
Using relation \rf{manus-17082023-04} and \rf{manus-17082023-30}, we find the $R^a$-transformations of the field strengths,
\beq
\label{manus-17082023-31} && \delta_{R^a} F_2^{bc} = 0\,,
\nonumber\\
&& \delta_{R^a} F_4^{bc} =  2 \eta^{ab} F_3^c - 2\eta^{ac} F_3^b  - 4 \partial^a  F_2^{bc} \,,
\nonumber\\
&& \delta_{R^a} F_6^{bc} =  4 \eta^{ab} F_5^c - 4\eta^{ac} F_5^b  - 4 \partial^a  F_4^{bc} \,,
\nonumber\\
&& \delta_{R^a} F_3^b  =  - 4 F_2^{ab}\,,
\nonumber\\
&& \delta_{R^a} F_5^b  =  - 2 F_4^{ab} - 2 \partial^a F_3^b\,.
\eeq

\noinbf{$F^3$-vertex of conformal YM field}.  We find a $F^3$-vertex denoted as $\LL_{F^3}$ which describes self-interaction of the conformal YM field. The vertex $\LL_{F^3}$ is of third order in the field strengths,
\beq
\label{manus-17082023-38} && \hspace{-1.5cm} g^{-1}\LL_{F^3} =  \LL^{(1)}  - \LL^{(2)}\,,
\nonumber\\
&& \LL^{(1)} \,\,\sim\,\, F_2^{ab} F_2^{bc} F_4^{ca}\,, \qquad  \LL^{(2)} \,\,\sim\,\, \half F_3^a F_2^{ab} F_3^b\,,
\eeq
where the field strengths are defined in \rf{manus-17082023-04}, while the sign $\,\,\sim\,\,$ is defined in \rf{manus-22082023-50}. The $g$ stands for a dimensionless coupling constant of the $F^3$-vertex.
Needles to say, the vertex $\LL_{F^3}$ \rf{manus-17082023-38} is invariant under gauge transformations \rf{manus-17082023-08} and conformal algebra transformations \rf{manus-22082023-33}, \rf{manus-17082023-31}. The vertex $\LL_{F^3}$ does not involve higher than third-order terms in the derivatives.

\noinbf{ Higher-derivative Lagrangian of conformal YM field}. Gauging away the Stueckelberg scalar fields $\phi_2$ and $\phi_4$ and ignoring the total derivatives, we represent the ordinary-derivative Lagrangian \rf{manus-17082023-03} as
\be \label{manus-17082023-70}
\LL_\smCYM \,\,\sim\,\, \phi_5^a D^bF_2^{ba}  +  \phi_3^a F_2^{ab}\phi_3^b - \frac{1}{4} F_4^{ab} F_4^{ab} - \phi_3^a\phi_5^a\,.
\ee
Using equations of motion for the auxiliary vector fields $\phi_5^a$ and $\phi_3^a$ obtained from Lagrangian \rf{manus-17082023-70}, we find the solution for the auxiliary vector fields $\phi_3^a$ and $\phi_5^a$, which we denote as $\phibf_3^a$ and $\phibf_5^a$ respectively,
\beq
\label{manus-17082023-71} && \phibf_3^a = D^b \Fbf_2^{ba}\,,
\nonumber\\
&& \phibf_5^a  = D^b\Fbf_4^{ba} + [\Fbf_2^{ab},\phibf_3^b]\,,\qquad \quad D^a \Fbf_n^{bc} := \partial^a \Fbf_n^{bc} + [\phi_1^a,\Fbf_n^{bc}]\,,
\eeq
where we use the notation
\beq
\label{manus-17082023-72} && \Fbf_2^{ab} = \partial^a \phi_1^b - \partial^b \phi_1^a +  [\phi_1^a,\phi_1^b]\,,
\nonumber\\
&& \Fbf_4^{ab} = D^a \phibf_3^b - D^b \phibf_3^a\,.
\eeq
Plugging the solutions $\phibf_3^a$ and $\phibf_5^a$ \rf{manus-17082023-71} and the field strengths \rf{manus-17082023-72} into the ordinary-derivative Lagrangian \rf{manus-17082023-70}, we get a higher-derivative (six-derivative) Lagrangian of conformal YM field, which we denote as $\LL_\smCYM^\smHder$,
\be \label{manus-17082023-73}
\LL_\smCYM^\smHder \,\,\sim\,\,  - \frac{1}{4} \Fbf_4^{ab} \Fbf_4^{ab} + \phibf_3^a \Fbf_2^{ab} \phibf_3^b\,,
\ee
where $\phibf_3^a$, $\Fbf_2^{ab}$, $\Fbf_4^{ab}$ are given in \rf{manus-17082023-71}, \rf{manus-17082023-72}. From \rf{manus-17082023-71}-\rf{manus-17082023-73}, we see that the higher-derivative Lagrangian is entirely expressed in terms of the generic (primary) field $\phi_1^a$ as it should be.

\noinbf{ Conformal YM field theory with two coupling constants}. Lagrangian with two coupling constant we want to discuss is described by two conformal invariants and is given by
\be \label{manus-17082023-87}
\LL_\tot = \LL_\smCYM + \LL_{F^3}\,,
\ee
where $\LL_\smCYM$, $\LL_{F^3}$ are given in \rf{manus-17082023-03}, \rf{manus-17082023-38}. Lagrangian $\LL_\tot$ \rf{manus-17082023-87} depends on the two conformal invariants given by $\LL_\smCYM$ and $\LL_{F^3}$. Note however that, for the YM field in $8d$, a number of conformal invariants is greater than two.%
\footnote{For example, for the YM field in $8d$, we can mention, besides $\LL_\smCYM$ and $\LL_{F^3}$, the conformal invariant $\LL_{F^4} \,\,\sim\,\, F_2^{ab} F_2^{bc}F_2^{ce} F_2^{ca}$ and many other similar $F^4$ conformal invariants built in terms of the field strength $F_2^{ab}$.
}
Namely, the conformal invariant $\LL_\smCYM$ involves terms of the second order and higher than the second order in the fields, while the conformal invariant $\LL_{F^3}$ involves terms of the third order and higher than the third order in the fields. All the remaining conformal invariants involve terms of higher than the third order in the fields. Those remaining conformal invariants are beyond of the scope of the present paper.

For the illustration purposes, consider a higher-derivative counterpart of the Lagrangian \rf{manus-17082023-87}. To this end, gauging away the Stueckelberg scalar fields $\phi_2$ and $\phi_4$ and using equations of motion for the vector fields $\phi_5^a$ and $\phi_3^a$ obtained from Lagrangian \rf{manus-17082023-87}, we find the solution for the vector fields $\phi_3^a$ and $\phi_5^a$, which we denote as $\phibf_{3\,\tot}^a$ and $\phibf_{5\,\tot}^a$ respectively,
\beq
\label{manus-17082023-88} && \phibf_{3\,\tot}^a = \phibf_3^a\,,
\nonumber\\
&& \phibf_{5\,\tot}^a  =  \phibf_5^a + g \phibf_{5\,F^3}^a\,,\qquad
\phibf_{5\,F^3}^a := D^b[\Fbf_2^{bc},\Fbf_2^{ca}] - \half [\Fbf_2^{ab},\phibf_3^b]\,,
\eeq
where the $\phibf_3^a$, $\phibf_5^a$ appearing in \rf{manus-17082023-88} take the same form as in \rf{manus-17082023-71}. Plugging the $\phibf_{3\,\tot}^a$, $\phibf_{5\,\tot}^a$ \rf{manus-17082023-88} into the Lagrangian $\LL_\tot$ \rf{manus-17082023-87}, we get the higher-derivative (six-derivative) Lagrangian,
\beq
\label{manus-17082023-90} && \LL_\tot^\smHder = \LL_\smCYM^\smHder + \LL_{F^3}^\smHder\,,
\nonumber\\
&&\hspace{1.4cm}  g^{-1}\LL_{F^3}^\smHder \,\,\sim\,\,  \Fbf_2^{ab} \Fbf_2^{bc} \Fbf_4^{ca} - \half \phibf_3^a \Fbf_2^{ab} \phibf_3^b\,,
\eeq
where $\LL_\smCYM^\smHder$ is given in \rf{manus-17082023-73}, while $\phibf_3^a$, $\Fbf_2^{ab}$, $\Fbf_4^{ab}$ are given in \rf{manus-17082023-71}, \rf{manus-17082023-72}. The following remarks are in order.

\noinbf{i)} The higher-derivative Lagrangian $\LL_\tot^\smHder$ given in \rf{manus-17082023-73}, \rf{manus-17082023-90} is completely expressed in terms of the generic (primary) field $\phi_1^a$. It other words, by using the gauge symmetries for the scalar fields and equations of motion for the auxiliary vector fields we can express our ordinary-derivative Lagrangian \rf{manus-17082023-87} entirely in terms of the field $\phi_1^a$.

\noinbf{ii})  Lagrangian $\LL_\smCYM^\smHder$ \rf{manus-17082023-73} involves six derivatives, while vertex $\LL_{F^3}^\smHder$ \rf{manus-17082023-90} involves five derivatives. To our knowledge, the Lagrangian $\LL_\tot^\smHder$ and the vertex $\LL_{F^3}^\smHder$ have not been discussed in the earlier literature. In other words, our approach allows us not only to derive the simple Lagrangian $\LL_\tot$ given in  \rf{manus-17082023-87} but also provides us the possibility to derive Lagrangian $\LL_\tot^\smHder$ \rf{manus-17082023-73} and vertex $\LL_{F^3}^\smHder$ \rf{manus-17082023-90} straightforwardly.

\noinbf{iii}) Lagrangian $\LL_\smCYM^\smHder$ \rf{manus-17082023-73} and vertex  $\LL_{F^3}^\smHder$ \rf{manus-17082023-90} are obviously invariant under the gauge transformations of the generic (primary) field $\phi_1^a$ given in \rf{manus-17082023-07}.

\noinbf{iv}) Lagrangian $\LL_\smCYM^\smHder$ \rf{manus-17082023-73} and vertex $\LL_{F^3}$ \rf{manus-17082023-90} are obviously invariant under the Poincar\'e and dilatation symmetries. To demonstrate  invariance of the Lagrangian $\LL_\smCYM^\smHder$ and the vertex $\LL_{F^3}^\smHder$ under the conformal boost transformations we use the following $R^a$-transformations of the various quantities appearing in \rf{manus-17082023-90}:

\beq
\label{manus-17082023-96} && \delta_{R^a} \Fbf_2^{bc} = 0\,,
\nonumber\\
&& \delta_{R^a} \phibf_3^b = - 4 \Fbf_2^{ab}\,,
\nonumber\\
&& \delta_{R^a} \Fbf_4^{bc} = 2\eta^{ab} \phibf_3^c - 2\eta^{ac} \phibf_3^b - 4 D^a \Fbf_2^{bc}\,.
\eeq
Using \rf{manus-17082023-96}, we find the relations (up to total derivative),
\be \label{manus-17082023-97}
\delta_{R^a} \LL_\smCYM^\smHder = 0 \,, \qquad \delta_{R^a} \LL_{F^3}^\smHder=0\,.
\ee
Relations \rf{manus-17082023-97} imply then that the Lagrangian $\LL_\smCYM^\smHder$ and the vertex $\LL_{F^3}^\smHder$ are invariant under conformal boost transformations.

\noinbf{v}) For the derivation of relations \rf{manus-17082023-96}, we use  \rf{manus-17082023-71}, \rf{manus-17082023-72} and note  that the $\Fbf_2^{bc}$, $\phibf_3^b$, and $\Fbf_4^{bc}$ are entirely expressed in terms of the generic field $\phi_1^b$. Therefore using the $R^a$-transformations of the field $\phi_1^b$ \rf{manus-17082023-30} and relations \rf{manus-17082023-71}, \rf{manus-17082023-72}, we find straightforwardly relations \rf{manus-17082023-96}.

\newsection{\large Conformal Yang-Mills field in ten dimensions } \label{ten}

{\bf Field content}. To discuss gauge invariant ordinary-derivative
formulation of the conformal YM field in $10d$ flat space we use four vector fields and
thee scalar fields:
\beq
\label{manus-18082023-01}   && \phi_1^a \qquad\quad \phi_3^a \qquad\quad \phi_5^a \qquad\quad \phi_7^a
\nonumber\\
&&
\\[-25pt]
&& \hspace{0.8cm} \phi_2 \qquad \quad \phi_4\qquad \quad \phi_6
\nonumber
\eeq
The fields $\phi_n^a$, $n=1,3,5,7$ are vector fields of the Lorentz algebra $so(9,1)$, while the fields $\phi_n$, $n=2,4,5$ are scalar fields of the $so(9,1)$.  Conformal dimensions of the fields in \rf{manus-18082023-01} are given by
\be \label{manus-18082023-02}
\Delta(\phi_{2n+1}^a) = 2n+1\,, \quad n=0,1,2,3\,; \hspace{1.5cm}  \Delta(\phi_{2n}) = 2n\,, \quad  n=1,2,3\,.
\ee
The field $\phi_1^a$ has the same conformal dimension as a conformal YM field  entering the higher-derivative approach. Therefore, in our approach, the field $\phi_1^a$ is identified with the conformal YM field, while the vector fields $\phi_{2n+1}^a$, $n=1,2,3$ turn out to be auxiliary fields. Below we show that the scalar fields $\phi_{2n}$ $n=1,2,3$ are realized as Stueckelberg fields.

\noinbf{Gauge invariant Lagrangian}. We use the sign $\,\,\sim\,\,$ given in \rf{manus-22082023-50}. The ordinary-derivative Lagrangian of the conformal YM field we found can then be presented in terms of field strengths as
\be \label{manus-18082023-03}
\LL_\smCYM  \,\,\sim\,\, -\half F_2^{ab} F_8^{ab} - \half F_4^{ab} F_6^{ab} -   F_3^a
F_7^a - \half F_5^a F_5^a\,,
\ee
where the field strengths are expressed in terms of the fields as
\beq
\label{manus-18082023-04} && F_2^{ab} = \partial^a \phi_1^b - \partial^b \phi_1^a +  [\phi_1^a,\phi_1^b]\,,
\nonumber\\
&& F_4^{ab} = D^a \phi_3^b - D^b \phi_3^a\,,
\nonumber\\
&& F_6^{ab}  = D^a \phi_5^b - D^b \phi_5^a +  \frac{4}{3}[\phi_3^a,\phi_3^b]\,,
\nonumber\\
&& F_8^{ab}  = D^a \phi_7^b - D^b \phi_7^a +  [\phi_3^a,\phi_5^b] + [\phi_5^a,\phi_3^b]\,,
\nonumber\\
&& F_3^a = \phi_3^a + D^a\phi_2\,,
\nonumber\\
&& F_5^a = \phi_5^a + D^a\phi_4  + \frac{2}{3} [\phi_3^a,\phi_2]\,,
\nonumber\\
&& F_7^a = \phi_7^a + D^a\phi_6  +  \frac{2}{3}  [\phi_3^a,\phi_4] + \frac{1}{3} [\phi_5^a,\phi_2]\,,
\eeq
while the realization of the covariant derivative $D^a$ on the auxiliary and Stueckelberg fields is given by
\be\label{manus-18082023-05}
D^a\phi_{2n+1}^b := \partial^a \phi_{2n+1}^b + [\phi_1^a,\phi_{2n+1}^b]\,, \qquad D^a\phi_{2n} := \partial^a \phi_{2n} + [\phi_1^a,\phi_{2n}]\,, \hspace{0.8cm} n = 1,2,3\,.
\ee
From \rf{manus-18082023-04}, we see that $F_2^{ab}$ is a field strength for the conformal YM field $\phi_1^a$, while, from \rf{manus-18082023-05}, we conclude that the  field $\phi_1^a$ is realized as the YM connection of the covariant derivative $D^a$.

\noinbf{Gauge transformations}. To discuss gauge symmetries of the Lagrangian \rf{manus-18082023-03} we introduce the following set of gauge parameters:
\be \label{manus-18082023-06}
\xi_0\,, \qquad \xi_2\,, \qquad \xi_4\,, \qquad \xi_6\,.
\ee
Gauge parameters in \rf{manus-18082023-06} are scalar fields of the Lorentz algebra $so(9,1)$. Gauge transformations we find take the following form:
\beq
\label{manus-18082023-07} && \delta_\xi \phi_1^b = D^b\xi_0\,,
\nonumber\\
&& \delta_\xi \phi_3^b = D^b\xi_2 + [\phi_3^b,\xi_0]\,,
\nonumber\\
&& \delta_\xi \phi_5^b = D^b\xi_4 + [\phi_5^b,\xi_0] + \frac{4}{3} [\phi_3^b,\xi_2]\,,
\nonumber\\
&& \delta_\xi \phi_7^b = D^b\xi_6 + [\phi_7^b,\xi_0] +  [\phi_5^b,\xi_2]+ [\phi_3^b,\xi_4]\,,
\nonumber\\
&& \delta_\xi \phi_2 = - \xi_2 + [\phi_2,\xi_0]\,,
\nonumber\\
&& \delta_\xi \phi_4 = - \xi_4 + [\phi_4,\xi_0] + \frac{2}{3} [\phi_2,\xi_2]\,,
\nonumber\\
&& \delta_\xi \phi_6 = - \xi_6 + [\phi_6,\xi_0] + \frac{2}{3} [\phi_4,\xi_2] + \frac{1}{3} [\phi_2,\xi_4]\,,
\eeq
where action of the covariant derivative $D^a$ on gauge parameters $\xi_{2n}$ \rf{manus-18082023-06} is given by
\be \label{manus-18082023-08}
D^a\xi_{2n} := \partial^a \xi_{2n} + [\phi_1^a,\xi_{2n}]\,.
\ee
The gauge transformations of the fields \rf{manus-18082023-07} and definition of field strengths \rf{manus-18082023-04} lead to the following gauge transformation of the field strengths:
\beq
\label{manus-18082023-09} && \delta_\xi F_2^{ab} = [F_2^{ab},\xi_0]\,,
\nonumber\\
&& \delta_\xi F_4^{ab} = [F_4^{ab},\xi_0] + [F_2^{ab},\xi_2]\,,
\nonumber\\
&& \delta_\xi F_6^{ab} = [F_6^{ab},\xi_0] + \frac{4}{3} [F_4^{ab},\xi_2]+ [F_2^{ab},\xi_4]\,,
\nonumber\\
&& \delta_\xi F_8^{ab} = [F_8^{ab},\xi_0] +  [F_6^{ab},\xi_2] +   [F_4^{ab},\xi_4] + [F_2^{ab},\xi_6]\,,
\nonumber\\
&& \delta_\xi F_3^a = [F_3^a,\xi_0]\,,
\nonumber\\
&& \delta_\xi F_5^a = [F_5^a,\xi_0] + \frac{2}{3}[F_3^a,\xi_2]\,,
\nonumber\\
&& \delta_\xi F_7^a = [F_7^a,\xi_0] + \frac{2}{3} [F_5^a,\xi_2]+ \frac{1}{3} [F_3^a,\xi_4]\,.
\eeq

\noinbf{Gauge algebra}. Let $\xi_n$ and $\eta_n$, $n=0,2,4,6$ be two sets of gauge transformation parameters. Using gauge transformations in \rf{manus-18082023-07}, we find the following commutators of two gauge transformations:
\beq
\label{manus-18082023-10} && \hspace{-0.8cm} [\delta_{\eta_0},\delta_{\xi_0}] =   \delta_{ [\eta_0,\xi_0] }\,,
\hspace{0.8cm} [\delta_{\eta_0},\delta_{\xi_2}] =   \delta_{ [\eta_0,\xi_2] }\,,
\hspace{0.6cm} [\delta_{\eta_0},\delta_{\xi_4}] =   \delta_{ [\xi_0,\xi_4] }\,,
\hspace{0.5cm} [\delta_{\eta_0},\delta_{\xi_6}] =   \delta_{ [\xi_0,\xi_6] }\,,\qquad
\nonumber\\[6pt]
&& \hspace{-0.8cm} [\delta_{\eta_2},\delta_{\xi_2}] =   \frac{4}{3} \delta_{ [\eta_2,\xi_2] }\,,
\hspace{0.5cm} [\delta_{\eta_2},\delta_{\xi_4}] = \delta_{ [\eta_2,\xi_4] }\,,
\hspace{0.7cm} [\delta_{\eta_2},\delta_{\xi_6}] = 0\,,
\nonumber\\[6pt]
&& \hspace{-0.8cm} [\delta_{\eta_4},\delta_{\xi_4}] = 0\,,
\hspace{1.6cm}  [\delta_{\eta_4},\delta_{\xi_6}] = 0\,,
\nonumber\\[6pt]
&& \hspace{-0.8cm} [\delta_{\eta_6},\delta_{\xi_6}] = 0\,.
\eeq
Using \rf{manus-18082023-10}, we verify that the Jacobi identities for three gauge transformations are satisfied.

\noinbf{$R^a$-transformations of fields and field strengths}. Transformations of the fields under action of the operator $R^a$ take the following form:
\beq
\label{manus-18082023-30} &&  \delta_{R^a} \phi_1^b  = 0\,,
\nonumber\\
&&  \delta_{R^a} \phi_3^b  =    - 2\eta^{ab} \phi_2 - 6\partial^a \phi_1^b\,,
\nonumber\\
&&  \delta_{R^a} \phi_5^b  =  - 4\eta^{ab} \phi_4 - 8 \partial^a \phi_3^b\,,
\nonumber\\
&&  \delta_{R^a} \phi_7^b  =  - 6\eta^{ab} \phi_6 - 6 \partial^a \phi_5^b\,,
\nonumber\\
&&  \delta_{R^a} \phi_2  = 6 \phi_1^a \,,
\nonumber\\
&& \delta_{R^a} \phi_4  = 4 \phi_3^a - 4\partial^a \phi_2\,,
\nonumber\\
&& \delta_{R^a} \phi_6  = 2 \phi_5^a - 4\partial^a \phi_4\,.
\eeq
Using relations \rf{manus-18082023-04}, \rf{manus-18082023-30}, we find the $R^a$-transformations of the field strengths,
\beq
\label{manus-18082023-31} && \delta_{R^a} F_2^{bc} = 0 \,,
\nonumber\\
&& \delta_{R^a} F_4^{bc} = 2\eta^{ab} F_3^c - 2\eta^{ac} F_3^b - 6\partial^a F_2^{bc} \,,
\nonumber\\
&& \delta_{R^a} F_6^{bc} = 4\eta^{ab} F_5^c - 4\eta^{ac} F_5^b - 8\partial^a F_4^{bc} \,,
\nonumber\\
&& \delta_{R^a} F_8^{bc} = 6\eta^{ab} F_7^c - 6\eta^{ac} F_7^b - 6\partial^a F_6^{bc} \,,
\nonumber\\
&& \delta_{R^a} F_3^b = - 6 F_2^{ab}\,,
\nonumber\\
&& \delta_{R^a} F_5^b = - 4 F_4^{ab} - 4\partial^a F_3^b\,,
\nonumber\\
&& \delta_{R^a} F_7^b = - 2 F_6^{ab} - 4\partial^a F_5^b\,.
\eeq

\noinbf{$F^3$-vertex of conformal YM field}. We find a $F^3$-vertex denoted as $\LL_{F^3}$ which describes self-interaction of the conformal YM field. The vertex $\LL_{F^3}$ is of third order in the field strengths,
\beq
\label{manus-18082023-50} && \hspace{-1.5cm} g^{-1} \LL_{F^3} = \LL^{(1)} -  \LL^{(2)}\,,
\nonumber\\
&&  \LL^{(1)} \,\,\sim\,\, F_2^{ab}F_2^{bc} F_6^{ca} +  \frac{4}{3} F_2^{ab} F_4^{bc} F_4^{ca}\,,
\nonumber\\
&& \LL^{(2)} \,\,\sim\,\,  \frac{2}{3}\big(  F_3^a F_2^{ab} F_5^b +   F_5^a F_2^{ab} F_3^b + \frac{2}{3} F_3^a F_4^{ab} F_3^b\big)\,,
\eeq
where the field strengths are defined in \rf{manus-18082023-04}, while the sign $\,\,\sim\,\,$ is defined in \rf{manus-22082023-50}. The $g$ stands for a dimensionless coupling constant.
Needles to say, the vertex $\LL_{F^3}$ \rf{manus-18082023-50} is invariant under gauge transformations \rf{manus-18082023-09} and conformal algebra transformations \rf{manus-22082023-33}, \rf{manus-18082023-31}. The vertex $\LL_{F^3}$ does not involve higher than third-order terms in the derivatives.

\noinbf{ Higher-derivative Lagrangian of conformal YM field}. Gauging away the Stueckelberg scalar fields $\phi_2$, $\phi_4$, and $\phi_6$ and ignoring total derivatives, we represent the ordinary-derivative Lagrangian \rf{manus-18082023-03} as
\be \label{manus-18082023-70}
\LL_\smCYM \,\,\sim\,\, \phi_7^a  D^bF_2^{ba} + \phi_5^a [F_2^{ab},\phi_3^b] + \phi_5^a D^bF_4^{ba} + \frac{4}{3} \phi_3^a F_4^{ab}\phi_3^b  -\phi_3^a \phi_7^a
- \half \phi_5^a \phi_5^a\,.
\ee
Using equations of motion for the auxiliary vector fields $\phi_7^a$, $\phi_5^a$, and $\phi_3^a$ obtained from Lagrangian \rf{manus-18082023-70}, we find the solution for the auxiliary vector fields $\phi_3^a$, $\phi_5^a$,  $\phi_7^a$, which we denote as $\phibf_3^a$, $\phibf_5^a$, and $\phibf_7^a$, respectively,
\beq
\label{manus-18082023-71} && \phibf_3^a = D^b\Fbf_2^{ba}\,,
\nonumber\\
&& \phibf_5^a = D^b\Fbf_4^{ba} + [\Fbf_2^{ab},\phibf_3^b]\,,
\nonumber\\
&& \phibf_7^a = D^b\Fbf_6^{ba} + [\Fbf_2^{ab},\phibf_5^b] + \frac{4}{3} [\Fbf_4^{ab},\phibf_3^b]\,,
\eeq
where we use the notation
\beq
\label{manus-18082023-72} && \Fbf_2^{ab} := \partial^a \phi_1^b - \partial^b \phi_1^a +  [\phi_1^a,\phi_1^b]\,,
\nonumber\\
&& \Fbf_4^{ab} := D^a \phibf_3^b - D^b \phibf_3^a\,,
\nonumber\\
&& \Fbf_6^{ab}  := D^a \phibf_5^b - D^b \phibf_5^a +  \frac{4}{3}[\phibf_3^a,\phibf_3^b]\,.
\eeq
Plugging the solutions $\phibf_3^a$, $\phibf_5^a$, and $\phibf_7^a$ \rf{manus-18082023-71} and the field strengths  \rf{manus-18082023-72} into the ordinary-derivative Lagrangian \rf{manus-18082023-70}, we get a higher-derivative (eight-derivative) Lagrangian, which we denote as $\LL_\smCYM^\smHder$,
\be \label{manus-18082023-73}
\LL_\smCYM^\smHder  \,\,\sim\,\, \half \phibf_5^a  \phibf_5^a + \frac{4}{3} \phibf_3^a \Fbf_4^{ab}\phibf_3^b\,.
\ee
From \rf{manus-18082023-71}-\rf{manus-18082023-73}, we see that the higher-derivative Lagrangian is entirely expressed in terms of the generic field $\phi_1^a$. Note that it is the use of the solutions $\phibf_3^a$,  $\phibf_5^a$, and the field strength $\Fbf_4^{ab}$ that allows us to write the concise expression for the higher-derivative Lagrangian $\LL_\smCYM^\smHder$ \rf{manus-18082023-73}. Note also that the $\phibf_3^a$,  $\phibf_5^a$, and $\Fbf_4^{ab}$ are naturally occurring in the framework of our approach.

\noinbf{ Conformal YM field theory with two coupling constants}. Lagrangian we want to discuss is described by two conformal invariants and is given by
\be \label{manus-18082023-87}
\LL_\tot = \LL_\smCYM + \LL_{F^3}\,,
\ee
where $\LL_\smCYM$, $\LL_{F^3}$ are given in \rf{manus-18082023-03}, \rf{manus-18082023-50}. Lagrangian $\LL_\tot$ \rf{manus-18082023-87} depends on the two conformal invariants given by $\LL_\smCYM$ and $\LL_{F^3}$. Note however that, for the YM field in $10d$, a number of conformal invariants is greater than two.%
\footnote{For example, for the YM field in $10d$, we can mention, besides $\LL_\smCYM$ and $\LL_{F^3}$, the conformal invariant $\LL_{F^5} \,\,\sim\,\, F_2^{ab} F_2^{bc}F_2^{ce} F_2^{cf}F_2^{fa}$ and many other similar $F^5$ conformal invariants built in terms of the field strength $F_2^{ab}$.
}
Namely, the conformal invariant $\LL_\smCYM$ involves terms of the second order and higher than the second order in the fields, while the conformal invariant $\LL_{F^3}$ involves terms of the third order and higher than the third order in the fields. All the remaining conformal invariants involve terms of higher than the third order in the fields. Those remaining conformal invariants are not considered in the present paper.

For the illustration purposes, consider a higher-derivative cousin of the Lagrangian \rf{manus-18082023-87}. To this end, gauging away the Stueckelberg scalar fields $\phi_2$, $\phi_4$, $\phi_6$ and using equations of motion for the auxiliary vector fields $\phi_7^a$, $\phi_5^a$, and $\phi_3^a$ obtained from Lagrangian \rf{manus-18082023-87}, we find the solution for the auxiliary vector fields $\phi_3^a$, $\phi_5^a$ and $\phi_7^a$, which we denote as $\phibf_{3\,\tot}^a$, $\phibf_{5\,\tot}^a$, and $\phibf_{7\,\tot}^a$ respectively,
\beq
\label{manus-18082023-88} && \phibf_{3\,\tot}^a = \phibf_3^a \,,
\nonumber\\
&&  \phibf_{5\,\tot}^a = \phibf_5^a  + g \phibf_{5\, F^3}^a\,,\qquad \phibf_{5\, F^3}^a := D^b[\Fbf_2^{be},\Fbf_2^{ea}] - \frac{2}{3} [\Fbf_2^{ab},\phibf_3^b]\,,\qquad\qquad
\nonumber\\
&& \phibf_{7\,\tot}^a = \phibf_7^a  + g \phibf_{7\, F^3}^a\,,
\eeq
where solutions $\phibf_3^a$, $\phibf_5^a$, and $\phibf_7^a$ \rf{manus-18082023-88} take the same form as in \rf{manus-18082023-71}. Note that, upon plugging  $\phibf_{3\,\tot}^a$ \rf{manus-18082023-88} into Lagrangian $\LL_\tot$ \rf{manus-18082023-87}, the $\phibf_{7\,\tot}^a$ does not contribute to Lagrangian \rf{manus-18082023-87}. Therefore we skip the details of $\phibf_{7\, \tot}^a$ \rf{manus-18082023-88} which is entirely expressed in terms of the field $\phi_1^a$.  From \rf{manus-18082023-71}, \rf{manus-18082023-88}, we see that all auxiliary vector fields are also expressed in terms of the primary field $\phi_1^a$. Plugging  $\phibf_{3\,\tot}^a$, $\phibf_{5\,\tot}^a$ \rf{manus-18082023-88} into Lagrangian $\LL_\tot$ \rf{manus-18082023-87}, we get the higher-derivative (eight-derivative) Lagrangian given by
\beq
\label{manus-18082023-89} && \LL_\tot^\smHder  \,\,\sim\,\, \half \phibf_{5\,\tot}^a  \phibf_{5\,\tot}^a + \frac{4}{3} \phibf_3^a \Fbf_4^{ab} \phibf_3^b
\nonumber\\
&& \hspace{1cm} +\,\,  \frac{2g}{3} \Big( [\Fbf_2^{bc}, \Fbf_2^{ce}][\phibf_3^e, \phibf_3^b]  +  \Fbf_2^{bc} [\Fbf_4^{ce}, \Fbf_4^{eb}] - \frac{1}{3} \Fbf_4^{bc} [\phibf_3^c, \phibf_3^b] \Big)\,,
\eeq
where the $\Fbf_4^{ab}$ is given in \rf{manus-18082023-72}. In order to show a dependence of the higher-derivative Lagrangian $\LL_\tot^\smHder$ \rf{manus-18082023-89} on the constant $g$, we plug the expression for $\phibf_{5\,\tot}^a$ \rf{manus-18082023-88} into \rf{manus-18082023-89} and represent $\LL_\tot^\smHder$ as
\beq
\label{manus-18082023-90} && \hspace{-1.4cm} \LL_\tot^\smHder = \LL_\smCYM^\smHder + g \LL_g^\smHder + g^2 \LL_{gg}^\smHder\,,
\nonumber\\
&& \LL_g^\smHder \,\,\sim\,\, \phibf_5^a \phibf_{5\, \Fbf^3}^a  + \frac{2}{3} \Big( [\Fbf_2^{bc}, \Fbf_2^{ce}][\phibf_3^e,\phibf_3^b]  +  \Fbf_2^{bc}[\Fbf_4^{ce},\Fbf_4^{eb}] - \frac{1}{3} \Fbf_4^{bc}[\phibf_3^c,\phibf_3^b] \Big)\,, \qquad
\nonumber\\
&& \LL_{gg}^\smHder \,\,\sim\,\, \half \phibf_{5\,F^3}^a\phibf_{5\,F^3}^a\,,
\eeq
where $\LL_\smCYM^\smHder$ is defined in \rf{manus-18082023-73},  while $\phibf_3^a$, $\phibf_5^a$, $\Fbf_2^{ab}$, $\Fbf_4^{ab}$, and $\phibf_{5\, F^3}^a$ are given in \rf{manus-18082023-71}, \rf{manus-18082023-72}, \rf{manus-18082023-88}.
Note that the vertex $\LL_{gg}^\smHder$ \rf{manus-18082023-90} does not involve terms of the second order and third order in the fields. The following remarks are in order.

\noinbf{i}) Lagrangian $\LL_\smCYM^\smHder$ \rf{manus-18082023-73} involves eight derivatives, while vertices $\LL_g^\smHder$ and $\LL_{gg}^\smHder$ \rf{manus-18082023-90} involve seven and six derivatives respectively.
To our knowledge, the Lagrangian $\LL_\smCYM^\smHder$ and the vertices $\LL_g^\smHder$, $\LL_{gg}^\smHder$  have not been discussed in literature. Thus our approach allows us not only to get the simple representation for Lagrangian $\LL_\tot$ \rf{manus-18082023-87} but also provides us the possibility to derive the  higher-derivative Lagrangian $\LL_\smCYM^\smHder$ \rf{manus-18082023-73} and the vertices $\LL_g^\smHder$, $\LL_{gg}^\smHder$ \rf{manus-18082023-90}.

\noinbf{ii)} The higher-derivative Lagrangian $\LL_\tot^\smHder$ \rf{manus-18082023-90} is completely expressed in terms of the field $\phi_1^a$.

\noinbf{iii}) The Lagrangian $\LL_\smCYM^\smHder$ and the vertices $\LL_g^\smHder$, $\LL_{gg}^\smHder$ \rf{manus-18082023-90} are obviously invariant under gauge transformations of the generic (primary) field $\phi_1^a$ given in \rf{manus-18082023-07}.

\noinbf{iv}) The Lagrangian $\LL_\smCYM^\smHder$ and the vertices $\LL_g^\smHder$, $\LL_{gg}^\smHder$ are obviously invariant under the Poincar\'e and dilatation symmetries. We now demonstrate invariance of the Lagrangian $\LL_\smCYM^\smHder$ and the vertices $\LL_g^\smHder$, $\LL_{gg}^\smHder$ under the conformal boost transformations. To this end we note the $R^a$-transformations of the various quantities appearing in \rf{manus-18082023-90}:
\beq
&& \delta_{R^a} \Fbf_2^{bc} = 0\,,
\nonumber\\
&& \delta_{R^a} \phibf_3^b = - 6 \Fbf_2^{ab}\,,
\nonumber\\
&& \delta_{R^a} \Fbf_4^{bc} = 2\eta^{ab} \phibf_3^c - 2\eta^{ac} \phibf_3^b - 6 D^a \Fbf_2^{bc}\,,
\nonumber\\
&& \delta_{R^a} \phibf_5^b = - 4 \Fbf_4^{ab} - 4 D^a \phibf_3^b\,,
\nonumber\\
\label{manus-18082023-96} && \delta_{R^a} \phibf_{5\, F^3}^b = 0\,.
\eeq
Using \rf{manus-18082023-96}, we find the relations (up to total derivative),
\be \label{manus-18082023-97}
\delta_{R^a} \LL_\smCYM^\smHder = 0 \,, \qquad \delta_{R^a} \LL_g^\smHder=0\,, \qquad \delta_{R^a} \LL_{gg}^\smHder =0\,.
\ee
Relations \rf{manus-18082023-97} imply that the Lagrangian $\LL_\smCYM^\smHder$ and the vertices $\LL_g^\smHder$, $\LL_{gg}^\smHder$ are indeed invariant under the conformal boost  transformations.

\noinbf{v}) Finally, we comment on the derivation of the $R^a$-transformations in \rf{manus-18082023-96}. From \rf{manus-18082023-71}, \rf{manus-18082023-72}, and \rf{manus-18082023-88}, we note that the solutions $\phibf_3^b$, $\phibf_5^b$, $\phibf_{5\,F^3}^b$ and the field strengths $\Fbf_2^{bc}$, $\Fbf_4^{bc}$,  are entirely expressed in terms of the field $\phi_1^b$. Therefore using the $R^a$-transformations of the field $\phi_1^b$ in \rf{manus-18082023-30} and relations \rf{manus-18082023-71}, \rf{manus-18082023-72}, \rf{manus-18082023-88}, we find straightforwardly the relations given in \rf{manus-18082023-96}.

\newsection{\large Conformal Yang-Mills field in arbitrary even dimensions } \label{arb}

{\bf Field content}. To discuss gauge invariant ordinary-derivative
formulation of the conformal YM field in $R^{d-1,1}$ space, $d$-even, we use the following set of  vector and scalar fields:
\be \label{manus-19082023-01}
\phi_{2n+1}^a \quad n=0,1,\ldots,\frac{d-4}{2};\hspace{1.5cm}  \phi_{2n} \quad n=1,2,\ldots,\frac{d-4}{2}\,.
\ee
The fields $\phi_{2n+1}^a$ in \rf{manus-19082023-01} are vector fields of the Lorentz algebra $so(d-1,1)$, while the fields $\phi_{2n}$ in \rf{manus-19082023-01} are scalar fields of the $so(d-1,1)$.  Conformal dimensions of the fields in \rf{manus-19082023-01} are given by
\be \label{manus-19082023-02}
\Delta(\phi_{2n+1}^a) = 2n+1\,, \hspace{2cm} \Delta(\phi_{2n}) = 2n\,.
\ee
The field $\phi_1^a$ has the same conformal dimension as the conformal YM field  entering the higher-derivative approach. Therefore, in our approach, the field $\phi_1^a$ is identified with the conformal YM field, while the vector fields $\phi_{2n+1}^a$, $n=1,2,\ldots, \half(d-4)$, turn out to be auxiliary fields. Below we show that the scalar fields $\phi_{2n}$, $n=1,2,\ldots, \half(d-4)$, are realized as Stueckelberg fields in our approach.

\noinbf{Gauge invariant Lagrangian}. We use the sign $\,\,\sim\,\,$ defined in \rf{manus-22082023-50}. The ordinary-derivative Lagrangian of the conformal YM field we found can then be presented in terms of field strengths as
\be \label{manus-19082023-03}
\LL_\smCYM  \,\,\sim\,\, - \frac{1}{4} \sum_{m,n=0,1,\ldots,\frac{d-4}{2} \atop m+n=\frac{d-4}{2}} F_{2m+2}^{bc}  F_{2n+2}^{bc}- \half \sum_{m,n=1,2,\ldots,\frac{d-4}{2}\atop m+n=\frac{d-2}{2}} F_{2m+1}^b  F_{2n+1}^b\,,
\ee
where a set of field strengths entering the Lagrangian \rf{manus-19082023-03} is given by
\be \label{manus-19082023-04}
F_{2n+2}^{bc}\,, \quad n=0,1,\ldots,\frac{d-4}{2}\,;\hspace{2cm}
F_{2n+1}^b\,, \quad n=1,2,\ldots,\frac{d-4}{2}\,.
\ee
Explicit expressions for the field strengths \rf{manus-19082023-04} in terms of the fields \rf{manus-19082023-01} are found to be
\beq
\label{manus-19082023-05} && F_2^{bc} = \partial^b\phi_1^c - \partial^c \phi_1^b +  [\phi_1^b,\phi_1^c]\,,
\nonumber\\[8pt]
&& F_{2n+2}^{bc} = D^b\phi_{2n+1}^c - D^c \phi_{2n+1}^b + \sum_{i=1,2,\ldots,n-1} \alpha_{n,i} [\phi_{2i+1}^b,\phi_{2n-2i+1}^c]\,,
\nonumber\\
&& F_{2n+1}^b = \phi_{2n+1}^b + D^b \phi_{2n} + \sum_{i=1,2,\ldots,n-1} \beta_{n,i} [\phi_{2i+1}^b,\phi_{2n-2i}]\,,
\nonumber\\
&& \hspace{1.6cm} n= 1,2,\ldots,\frac{d-4}{2}\,,
\\
\label{manus-19082023-06} && \hspace{1.6cm} \alpha_{n,i} = \frac{n!}{i!(n-i)!} \frac{(\frac{d-4}{2}-n+i)! (\frac{d-4}{2}-i)!}{(\frac{d-4}{2})! (\frac{d-4}{2}-n)!} \,,
\nonumber\\
&& \hspace{1.6cm} \beta_{n,i} = \frac{n-i}{n}\alpha_{n,i}\,,
\eeq
where action of the covariant derivative $D^a$ on the auxiliary and Stueckelberg fields is given by
\beq
\label{manus-19082023-07} && D^a\phi_{2n+1}^b := \partial^a \phi_{2n+1}^b + [\phi_1^a,\phi_{2n+1}^b]\,, \qquad n = 1,2,\ldots,\frac{d-4}{2}\,;
\nonumber\\
&& D^a\phi_{2n} := \partial^a \phi_{2n} + [\phi_1^a,\phi_{2n}]\,, \hspace{2cm} n = 1, 2,\ldots,\frac{d-4}{2}\,.
\eeq
From \rf{manus-19082023-05}, we see that $F_2^{ab}$ is a standard field strength for the conformal YM field $\phi_1^a$, while, from \rf{manus-19082023-07},
we learn that the  field $\phi_1^a$ is realized as the YM connection of the covariant derivative $D^a$. As an side remark, we note that the field strengths \rf{manus-19082023-05} can alternatively be represented as
\beq
\label{manus-19082023-08} && F_{2n+2}^{bc} = \partial^b\phi_{2n+1}^c - \partial^c \phi_{2n+1}^b + \Gamma_{2n+2}^{bc}\,, \hspace{1cm} n= 0,1,\ldots,\frac{d-4}{2}\,;
\nonumber\\
&& F_{2n+1}^b = \phi_{2n+1}^b + \partial^b \phi_{2n} + \Gamma_{2n+1}^b\,, \hspace{1.8cm} n= 1,2,\ldots,\frac{d-4}{2}\,;
\\
&& \label{manus-19082023-09} \hspace{1.3cm} \Gamma_{2n+2}^{bc} := \sum_{i=0,1,\ldots,n} \alpha_{n,i} [\phi_{2i+1}^b,\phi_{2n-2i+1}^c]\,,
\nonumber\\
&&  \hspace{1.3cm} \Gamma_{2n+1}^b :=   \sum_{i=0,1,\ldots,n-1} \beta_{n,i} [\phi_{2i+1}^b,\phi_{2n-2i}]\,.
\eeq
Expressions for the field strengths given in \rf{manus-19082023-08}, \rf{manus-19082023-09} turn out to be more convenient for the study of conformal and gauge transformations of the field strengths presented in Appendices A,B.
In \rf{manus-19082023-03}, in the expression for the first sum, we note the restriction $m+n = \frac{d-4}{2}$,
while, in the expression for the second sum, we note the restriction $m + n = \frac{d-2}{2}$. These restrictions are simply obtained by using the requirement that the conformal dimension of the Lagrangian \rf{manus-19082023-03} should be equal to $d$ and taking into account the conformal dimensions of field strengths given in \rf{manus-22082023-31}. The Lagrangian \rf{manus-19082023-03} is invariant under gauge transformations and conformal algebra transformations. Lagrangian \rf{manus-19082023-03} is obviously invariant under Poincar\'e algebra and dilatation symmetries.
The Lagrangian \rf{manus-19082023-03} is easily fixed by requiring the Lagrangian to be invariant under conformal boost transformations which are discussed below.

\noinbf{Gauge transformations}. To discuss gauge symmetries of the Lagrangian \rf{manus-19082023-03} we introduce the following set of gauge parameters:
\be \label{manus-19082023-10}
\xi_{2n}\,, \qquad n=0,1,\ldots, \frac{d-4}{2}\,.
\ee
Gauge parameters in \rf{manus-19082023-10} are scalar fields of the Lorentz algebra $so(d-1,1)$. Conformal dimensions of the gauge parameters are given by $\Delta(\xi_{2n})=2n$. Gauge transformations of fields \rf{manus-19082023-01} take the following form:
\beq
\label{manus-19082023-11} && \delta_\xi \phi_{2n+1}^b = D^b \xi_{2n} +   \sum_{i=0,1,\ldots,n-1} \alpha_{n,i} [\phi_{2n+1-2i}^b,\xi_{2i}]\,,\hspace{1cm}  n= 0,1,\ldots,\frac{d-4}{2}\,;\qquad
\nonumber\\
&& \delta_\xi \phi_{2n} = - \xi_{2n}   + \sum_{i=0,1,\ldots,n-1} \beta_{n,i} [\phi_{2n-2i},\xi_{2i}]\,, \hspace{2cm}   n = 1,2,\ldots,\frac{d-4}{2}\,;
\eeq
where an action of the covariant derivative $D^a$ on the gauge parameters $\xi_{2n}$ \rf{manus-19082023-10} is given by
\be \label{manus-19082023-11-a}
D^a\xi_{2n} := \partial^a \xi_{2n} + [\phi_1^a,\xi_{2n}]\,.
\ee
For $n=0$, formula \rf{manus-19082023-11} implies $\delta\phi_1^b = D^b\xi_0$. From \rf{manus-19082023-11}, we see that the one-derivative contributions to the gauge transformations of the vector fields take the form of
the standard gradient gauge transformations of vector fields, while all scalar fields transform as the Stueckelberg fields.

Using \rf{manus-19082023-05}, we find that the gauge transformations of the fields \rf{manus-19082023-11} lead to following gauge transformations of the field strengths:
\beq
\label{manus-19082023-12} && \delta_\xi F_{2n+2}^{bc} =  \sum_{i=0,1,\ldots,n} \alpha_{n,i} [F_{2n+2-2i}^{bc},\xi_{2i}]\,,\hspace{1.4cm}  n = 0,1,\ldots,\frac{d-4}{2}\,;\qquad
\nonumber\\
&& \delta_\xi F_{2n+1}^b =  \sum_{i=0,1,\ldots,n-1} \beta_{n,i} [F_{2n+1-2i}^b,\xi_{2i}]\,, \hspace{1cm} n= 1,2,\ldots,\frac{d-4}{2}\,;
\eeq
where the coefficients $\alpha_{n,i}$, $\beta_{n,i}$ are given in \rf{manus-19082023-06}.

The linearized part of the gauge transformations of the fields in \rf{manus-19082023-11} was obtained in Ref.\cite{Metsaev:2007fq}. The non-abelian contributions appearing in the gauge transformations of the fields \rf{manus-19082023-11} as well as gauge transformations of the fields strengths \rf{manus-19082023-12} are obtained in this paper. For the derivation of \rf{manus-19082023-11} and \rf{manus-19082023-12}, see Appendix B.

\noinbf{Gauge algebra}. Let $\xi_{2n}$ and $\eta_{2n}$, $n=0,1,\ldots,\frac{d-4}{2}$, be two sets of gauge parameters. Using the gauge transformations in \rf{manus-19082023-11}, we find the following commutators of two gauge transformations:
\beq
\label{manus-19082023-20} && [\delta_{\eta_{2i}},\delta_{\xi_{2j}}] = \delta_{[\eta_{2i}\star\,\xi_{2j}]}\,,
\\
\label{manus-19082023-21} && [\eta_{2i}\star\xi_{2j}] : = \alpha_{i+j,j}[\eta_{2i},\xi_{2j}]\,,
\eeq
where a star commutator appearing on r.h.s in \rf{manus-19082023-20} is defined in terms of the usual commutator in \rf{manus-19082023-21}, while the coefficients $\alpha_{m,n}$ entering the star commutator are given in \rf{manus-19082023-06}. Using \rf{manus-19082023-20}, it is easy to see that the Jacobi identity for three gauge transformations amounts to the following star Jacobi identity:
\be \label{manus-19082023-22}
[[\eta_{2i}\star\xi_{2j}] \,\star\, \chi_{2k}] + [[\chi_{2k}\star\eta_{2i}]\star\xi_{2j}]
+ [[\xi_{2j}\star\chi_{2k}]\star\eta_{2i}] = 0 \,.
\ee
Using \rf{manus-19082023-21}, we verify that the star Jacobi identity \rf{manus-19082023-22} is indeed satisfied by virtue of the Jacobi identity for the usual commutator and the following property of the coefficients $\alpha_{m,n}$ \rf{manus-19082023-06}:
\be \label{manus-19082023-23}
\alpha_{i+j,j}\alpha_{i+j+k,k} \ \hbox{ is symmetric upon all permutations of } i,j,k\,.
\ee

\noinbf{$R^a$-transformations of fields and field strengths}. Transformations of the fields \rf{manus-19082023-01} under action of the operator $R^a$ were found in Ref.\cite{Metsaev:2007fq} and they are given by
{\small
\beq
\label{manus-19082023-30} && \hspace{-1.3cm} \delta_{R^a} \phi_{2n+1}^b =  - 2n \eta^{ab}
\phi_{2n} - n(d-2-2n)\partial^a \phi_{2n-1}^b\,,\hspace{2.5cm} n=0,1,\ldots,\frac{d-4}{2}\,,\qquad
\nonumber\\
&& \hspace{-1.3cm} \delta_{R^a} \phi_{2n} =  (d-2-2n)
\phi_{2n-1}^a -  (n-1)(d-2-2n)\partial^a \phi_{2n-2}\,,  \qquad n=1,2,\ldots,\frac{d-4}{2}\,.\qquad
\eeq
}
Using relations \rf{manus-19082023-05}, \rf{manus-19082023-30}, we now find the $R^a$-transformations of the field strengths \rf{manus-19082023-04},
{\small
\beq
\label{manus-19082023-31} && \hspace{-1.3cm} \delta_{R^a} F_{2n+2}^{bc} =  2n\big(\eta^{ab} F_{2n+1}^c -\eta^{ac} F_{2n+1}^b\big) -  n (d-2-2n)\partial^a F_{2n}^{bc}\,,\qquad n=0,1,\ldots,\frac{d-4}{2}\,,\qquad
\nonumber\\
&&  \hspace{-1.3cm} \delta_{R^a} F_{2n+1}^b = - (d-2-2n) F_{2n}^{ab} - (n-1)(d-2-2n)\partial^a F_{2n-1}^b\,, \qquad n=1,2,\ldots,\frac{d-4}{2}\,.\qquad
\eeq
}
Transformation rules \rf{manus-22082023-33}, \rf{manus-22082023-34} and explicit expression for the $R^a$- transformations given in \rf{manus-19082023-31} provides us the complete description of the conformal boost transformations for field strengths. Using these transformations rules it is easy to demonstrate that the only ordinary-derivative Lagrangian which is invariant under conformal algebra transformations is given by \rf{manus-19082023-03}.

\noinbf{$F^3$-vertex of conformal YM field}. We find $F^3$-vertex denoted as $\LL_{F^3}$ which describes self-interaction of the conformal YM field. The vertex $\LL_{F^3}$ is of third order in the field strengths,
\beq
\label{manus-19082023-50} && \hspace{-1.3cm} g^{-1} \LL_{F^3} = \LL^{(1)} - \LL^{(2)}\,,
\nonumber\\[9pt]
&& \LL^{(1)} \,\,\sim\,\, \sum_{n_1,n_2,n_3 =0,1,\ldots, \frac{d-6}{2}\atop n_1 + n_2 + n_3 = \frac{d-6}{2} } c_{n_1,n_2,n_3}^{(1)} F_{2n_1+2}^{ab} F_{2n_2+2}^{bc} F_{2n_3+2}^{ca}\,,
\nonumber\\
&& \LL^{(2)} \,\,\sim\,\, \sum_{ {n_1,n_3 =1,2,\ldots, \frac{d-6}{2} \atop n_2 =0,1,\ldots, \frac{d-8}{2}} \atop n_1 + n_2 + n_3 = \frac{d-4}{2}   } c_{n_1,n_2,n_3}^{(2)} F_{2n_1+1}^a F_{2n_2+2}^{ab} F_{2n_3+1}^b\,,
\eeq
where sign $\,\,\sim\,\,$ is defined in \rf{manus-22082023-50}, while $g$ stands for a coupling constant. The field strengths are defined in \rf{manus-19082023-05}, while the coefficients $c_{n_1n_2n_3}^{(1),(2)}$ appearing in \rf{manus-19082023-50} are given by
\beq
\label{manus-19082023-51} && c_{n_1,n_2,n_3}^{(1)} := \frac{(\frac{d-4}{2}-n_1)! (\frac{d-4}{2}-n_2)! (\frac{d-4}{2}-n_3)!}{n_1! n_2! n_3!}c_0^{(1)}\,,
\nonumber\\
&& c_{n_1,n_2,n_3}^{(2)} := \frac{(\frac{d-4}{2}-n_1)! (\frac{d-4}{2}-n_2)! (\frac{d-4}{2}-n_3)!}{(n_1-1)! n_2! (n_3-1)!}c_0^{(2)}\,,
\nonumber\\
&& \hspace{1.8cm} c_0^{(1)} := \frac{1}{3(\frac{d-4}{2})(\frac{d-4}{2})!}\,, \qquad c_0^{(2)} :=  \frac{1}{(\frac{d-4}{2})(\frac{d-4}{2})!}\,.
\eeq
From \rf{manus-19082023-50}, we see that the vertex $\LL_{F^3}$ does not involve higher than third-order terms in the derivatives, while, from \rf{manus-19082023-51}, we see that the vertex $\LL_{F^3}$ is fixed uniquely up to one dimensionless coupling constant $g$. The vertex $\LL_{F^3}$ \rf{manus-19082023-50} is invariant under gauge transformations given in \rf{manus-19082023-12} and conformal algebra transformations described in Sec.\ref{not-conv} and \rf{manus-19082023-31}.  The conformal and gauge symmetries determine the vertex $\LL_{F^3}$ uniquely. In Appendix C, we outline a procedure of the derivation of the vertex $\LL_{F^3}$.

\noinbf{Interrelation between ordinary-derivative and higher-derivative approaches}. Here we demonstrate that our approach is equivalent to the standard higher-derivative approach.
Namely, we are going to demonstrate that all auxiliary fields can be integrated out at non-linear level leading just to a local higher-derivative action which is expressed only in terms of the field strength $F_2^{bc}$ and the covariant derivative $D^a$. To this end we note that, in our approach, the most general Lagrangian for the conformal YM field can be presented as
\beq
\label{manus-19102023-70} && \LL_\tot = \LL_\smCYM + \LL_\ext\,,
\\
\label{manus-19082023-71} && \LL_\ext = \LL_\ext(D^a,F^3,F^4,\ldots,F^{d/2})\,,
\eeq
where $\LL_\smCYM$ is given in \rf{manus-19082023-03}, while a local vertex $\LL_\ext$ depends on $F^n$ vertices, $n=3,4,\ldots,d/2$, and covariant derivative $D^a$. Explicit expression of the vertxe $\LL_\ext$ is not required for our discussion. All we need is the explicit expression for $\LL_\smCYM$ \rf{manus-19082023-03} and restrictions imposed by dilatation symmetry. Namely, gauging away all scalar fields, we note that equations of motion for the auxiliary fields $\phi_{d-3}^a$, $\phi_{d-5}^a$, \ldots, $\phi_3^a$ obtained from Lagrangian \rf{manus-19102023-70} take the following respective form:
\beq
\label{manus-19082023-72} && \phi_3^a = E_3^a(D^b, F_2^{bc})\,,
\\
\label{manus-19082023-73} && \phi_5^a = E_5^a(D^b, F_2^{bc},\phi_3^b)\,,
\\
\label{manus-19082023-74} && \phi_7^a = E_7^a(D^b, F_2^{bc},\phi_3^b,\phi_5^b)\,,
\\
\label{manus-19082023-75} && \phi_9^a = E_9^a(D^b, F_2^{bc},\phi_3^b,\phi_5^b,\phi_7^b)\,,
\\
&& \ldots\quad \ldots
\nonumber\\
\label{manus-19082023-76} && \phi_{d-3}^a = E_{d-3}^a(D^b, F_2^{bc},\phi_3^b,\ldots, \phi_{d-5}^b)\,,
\eeq
where $E_n^a$ are local polynomials of the field strength $F_2^{bc}$ and the auxiliary fields. Obviously the $E_n^a$ are uniquely fixed by the Lagrangian $\LL_\tot$.  Equations  \rf{manus-19082023-72}-\rf{manus-19082023-76} tell us that all auxiliary fields $\phi_3^a,\ldots, \phi_{d-3}^a$ and hence the Lagrangian $\LL_\tot$ can completely be expressed in terms of the field strength $F_2^{bc}$. For example, from equation \rf{manus-19082023-72}, we find $\phi_3^a$. Plugging such $\phi_3^a$ into equation \rf{manus-19082023-73}, we find $\phi_5^a$, and so on.

For the reader convenience, we make comment on equations \rf{manus-19082023-72}-\rf{manus-19082023-76}. The $\phi_n^a$-terms on the left hand side of equations \rf{manus-19082023-72}-\rf{manus-19082023-76} are obtained by varying the $F_m^a F_n^a$-terms \rf{manus-19082023-03} with respect to the auxiliary field $\phi_{d-n}^a$, while the expressions for the $E_n^a$ are obtained by varying the vertex $\LL_\ext$ and the $F_m^{bc} F_n^{bc}$-terms \rf{manus-19082023-03} with respect to the auxiliary field $\phi_{d-n}^a$. The triangle form of equations \rf{manus-19082023-72}-\rf{manus-19082023-76} can easily be understood by using the dilatation symmetry and the fact that the $E_n^a$ are local polynomials of the field strength $F_2^{bc}$ and the auxiliary fields.. For example, the auxiliary fields $\phi_5^a$,\ldots, $\phi_{d-3}^a$ cannot appear in equation \rf{manus-19082023-72}. If this was not true, then the dilatation symmetry would be broken.

As an side remark we note that the $E_3^a$ entering \rf{manus-19082023-72} can easily be found. Namely, we note that equation \rf{manus-19082023-72} is obtained by varying the Lagrangian $\LL$ with respect to the auxiliary field $\phi_{d-3}^a$. Taking into account that a conformal dimension of the Lagrangian $\LL_\tot$ should be equal to $d$, it is easy to understand that the auxiliary field $\phi_{d-3}^a$ enters only the Lagrangian $\LL_\smCYM$ and does not appear in the vertex $\LL_\ext$. Taking into account $\LL_\smCYM$ given in \rf{manus-19082023-03}, we then find
\be
E_3^a = D^b F_2^{ba}\,.
\ee

\newsection{\large Conclusions } \label{concl}

In this paper, we applied the ordinary-derivative formulation of conformal fields developed in Refs.\cite{Metsaev:2007fq,Metsaev:2007rw} to the study of interacting conformal YM field. We have shown that our approach provides us a possibility to build ordinary-derivative Lagrangian of the conformal YM field in arbitrary even dimensions in a rather straightforward way. Also we applied our approach for the building of the $F^3$-vertex of the conformal YM field. Our approach might have the following potentially interesting applications and generalizations.

\noinbf{i}) In Ref.\cite{Metsaev:2016rpa}, by using a light-cone approach, we build all cubic interaction vertices for conformal vector and scalar fields. We expect that our approach in this paper provides a possibility for the relatively simple generalization of the light-cone gauge vertices to Lorentz covariant and gauge invariant vertices to all orders in fields. It seems to be of some interest to apply our ordinary-derivative approach to the study of Lorentz covariant and gauge invariant conformal interactions between conformal YM vector field and matter fields. Discussion of conformal interactions between matter fields and conformal fields may be found, e.g., in Refs.\cite{Bekaert:2010ky,Manvelyan:2006bk,Kuzenko:2022hdv}.

\noinbf{ii}) As we have already noted, kinetic terms entering our ordinary-derivative Lagrangian formulation of the conformal fields turn out to be the conventional Klein-Gordon, Maxwell and Einstein-Hilbert kinetic terms. We think therefore that our approach will provide new interesting possibilities for the quantization of conformal fields and the computation of conformal fields scattering amplitudes. The recent intensive study of conformal fields scattering amplitudes may be found in Refs.\cite{Beccaria:2016syk,Joung:2015eny,Adamo:2018srx}.%
\footnote{ Study of conformal higher-spin scattering amplitudes by using twistors may be found in Ref.\cite{Adamo:2016ple}. Recent interesting applications of twistors to YM-like theories and higher-spin gravity may be found in Refs.\cite{Steinacker:2023zrb}-\cite{Basile:2022mif}.
}
The interesting discussion of hidden conformal symmetries of scattering amplitudes for gravitons in Einstein theory may be found in Ref.\cite{Loebbert:2018xce}, while the study of hidden conformal symmetries of higher-spin gravity may be found in Refs.\cite{Vasiliev:2007yc}.

\noinbf{iii}) Supersymmetric higher-derivative Lagrangian of the conformal YM field in $6d$  was obtained in Ref.\cite{Ivanov:2005qf}. In Ref.\cite{Metsaev:2012hr}, we developed the ordinary-derivative formulation of free fermionic conformal fields. We expect therefore that the formulation for the fermionic conformal fields obtained in Ref.\cite{Metsaev:2012hr} and our result in this paper will make it possible to develop an ordinary-derivative formulation of supersymmetric conformal YM field theories in a rather straightforward way.%
\footnote{ Various interesting alternative studies of fermionic fields may be found in Refs.\cite{Vasiliev:1987tk}-\cite{Lindwasser:2023zwo}. We mention also interesting Ref.\cite{Bergshoeff:1982az} devoted to the study of conformal supergravity in $10d$.
}

\noinbf{iv}) One-loop beta-function for conformal supersymmetric YM theory in $6d$ was computed by various methods in Refs.\cite{Buchbinder:2020tnc,Casarin:2019aqw}. It is desirable to understand about whether or not and in which way our ordinary-derivative approach might be helpful for the investigation of quantum properties of conformal supersymmetric YM theories.

\noinbf{v}) In Ref.\cite{Metsaev:2015yyv}, we developed the ordinary-derivative BRST-BV formulation of free arbitrary spin conformal fields. Generalization of the formulation in Ref.\cite{Metsaev:2015yyv} to the case of interacting conformal fields along the BRST-BV method in Refs.\cite{Dempster:2012vw,Metsaev:2012uy,Buchbinder:2021xbk} could be of some interest. Recent interesting study of the BRST-BV approach may be found in Refs.\cite{Reshetnyak:2023oyj}.

\medskip

\noinbf{Acknowledgment}. We thank the referee of Nucl. Phys. B for the suggestion to demonstrate an equivalence of the standard higher-derivative approach and our ordinary-derivative approach.

\setcounter{section}{0}\setcounter{subsection}{0}
\appendix{ \large Derivation of expressions for $\alpha_{n,i}$, $\beta_{n,i}$ \rf{manus-19082023-06} }

We start with the linearized gauge transformations of fields \rf{manus-22082023-02} found in Ref.\cite{Metsaev:2007fq},
\be \label{manus-28082023-01}
\delta^\lin \phi_{2n+1}^b = \partial^b \xi_{2n}\,,\hspace{1cm}  \delta^\lin \phi_{2n} = - \xi_{2n}\,,
\ee
and build the following linearized field strengths which are invariant under gauge transformations \rf{manus-28082023-01}:
\beq
\label{manus-28082023-02} && f_{2n+2}^{bc} = \partial^b\phi_{2n+1}^c - \partial^c \phi_{2n+1}^b  \,, \hspace{1cm} n= 0,1,\ldots,\frac{d-4}{2}\,,
\nonumber\\
&& f_{2n+1}^b = \phi_{2n+1}^b + \partial^b \phi_{2n}\,, \hspace{1.8cm} n= 1,2,\ldots,\frac{d-4}{2}\,.
\eeq
The $R^a$-transformations of the fields found in Ref.\cite{Metsaev:2007fq} are given in \rf{manus-19082023-30}. Using \rf{manus-22082023-07}, \rf{manus-22082023-08}, and \rf{manus-19082023-30}, we the following $R^a$-transformations of linearized field strengths \rf{manus-28082023-02}:
\beq
\label{manus-28082023-03}  && \delta_{R^a}  f_{2n+2}^{bc}  =   2n ( \eta^{ab} f_{2n+1}^c - \eta^{ac} f_{2n+1}^b )  - n(d-2-2n) \partial^a f_{2n}^{bc}\,,
\nonumber\\
&&\delta_{R^a} f_{2n+1}^b = - (d-2-2n) f_{2n}^{ab} - (n-1)(d-2-2n)\partial^a f_{2n-1}^b\,.
\eeq
We now assume that full field strengths denoted as $F_{2n+2}^{bc}$ and $F_{2n+1}^b$ have the same $R^a$- transformations as their respective linearized counterparts $f_{2n+2}^{bc}$ and $f_{2n+1}^b$ in \rf{manus-28082023-03}. Namely, we assume that $R^a$-transformations of the field strengths $F_{2n+2}^{bc}$ and $F_{2n+1}^b$ take the form as in \rf{manus-19082023-31}. We then verify that expressions  \rf{manus-19082023-08} and the $R^a$-transformations of the field strengths \rf{manus-19082023-31} require that the $R^a$-transformations of the $\Gamma_{2n+2}^{bc}$, $\Gamma_{2n+1}^b$ should take the following form:
\beq
\label{manus-28082023-04}  && \delta_{R^a}  \Gamma_{2n+2}^{bc}  =   2n ( \eta^{ab} \Gamma_{2n+1}^c - \eta^{ac} \Gamma_{2n+1}^b )  - n(d-2-2n) \partial^a \Gamma_{2n}^{bc}\,,
\\
\label{manus-28082023-05}  &&\delta_{R^a} \Gamma_{2n+1}^b = - (d-2-2n) \Gamma_{2n}^{ab} - (n-1)(d-2-2n)\partial^a \Gamma_{2n-1}^b\,.
\eeq
Now we are ready to formulate our

\noinbf{Statement}. Let the $R^a$-transformations of the fields be given by  \rf{manus-19082023-30}, while expressions for the $\Gamma_{2n+2}^{bc}$ and $\Gamma_{2n+1}^b$ be given by \rf{manus-19082023-09}, where $\alpha_{n,i}$ and $\beta_{n,i}$ are unknown coefficients. Then requiring that the $R^a$-transformations of the $\Gamma_{2n+2}^{bc}$ and $\Gamma_{2n+1}^b$ be given by \rf{manus-28082023-04} and \rf{manus-28082023-05},  we find that the coefficients $\alpha_{n,i}$ and $\beta_{n,i}$ are fixed uniquely up to one coupling constant and they are given in \rf{manus-19082023-06}.

\noinbf{Proof of Statement}. Firstly, by using transformation rules  \rf{manus-19082023-30} and expressions \rf{manus-19082023-09}, we verify that  transformation rules \rf{manus-28082023-04} amount to the following  equations for $\alpha_{n,i}$ and $\beta_{n,i}$:
\beq
\label{manus-28082023-06} && (i+1)(d-4-2i) \alpha_{n,i+1} = (n-i)(d-2-2n+2i)\alpha_{n,i}\,,
\\
\label{manus-28082023-07} && (n-i)(d-2-2n+2i)\alpha_{n,i} = n(d-2-2n) \alpha_{n-1,i}\,,
\\
\label{manus-28082023-08} && \alpha_{n,i} (2n-2i) = 2n \beta_{n,i}\,,
\eeq
while transformation rules \rf{manus-28082023-05} amount to the following  equations for $\alpha_{n,i}$ and $\beta_{n,i}$:
\beq
\label{manus-28082023-09} && (i+1)(d-4-2i)\beta_{n,i+1} = (n-1-i)(d-2-2n+2i)\beta_{n,i}\,,
\\
\label{manus-28082023-10} && (i+1)(d-4-2i)\beta_{n,i+1} = (n-1)(d-2-2n)\beta_{n-1,i}\,,
\\
\label{manus-28082023-11} && (d-4-2i)\beta_{n,n-1-i} = (d-2-2n)\alpha_{n-1,i}\,.
\eeq
Secondly, we find that the solutions to equations \rf{manus-28082023-06}, \rf{manus-28082023-07} and \rf{manus-28082023-08}, \rf{manus-28082023-09} are given by the following respective expressions for $\alpha_{n,i}$ and $\beta_{n,i}$:
\beq
\label{manus-28082023-20} && \alpha_{n,i} =  \frac{n!}{i!(n-i)!} \frac{(\frac{d-4}{2}-n+i)! (\frac{d-4}{2}-i)!}{(\frac{d-4}{2}-n)!} \alpha\,,
\\
\label{manus-28082023-21} && \beta_{n,i} = \frac{(n-1)!}{i!(n-1-i)!} \frac{(\frac{d-4}{2}-n+i)! (\frac{d-4}{2}-i)!}{ (\frac{d-4}{2}-n)!} \beta\,,
\eeq
where coefficients $\alpha$ and $\beta$ do not depend on $n$ and $i$.

Thirdly, we verify that the remaining equations \rf{manus-28082023-08} and \rf{manus-28082023-11} are satisfied provided $\alpha=\beta$. Using solution \rf{manus-28082023-20}, we find
\be \label{manus-28082023-22}
F_2^{ab} = \partial^a \phi_1^b - \partial^b \phi_1^a + \alpha_{0,0} [\phi_1^a,\phi_1^b]
\ee
Finally, using the normalization $\alpha_{0,0}=1$ which can be achieved by re-scaling of the fields, we get
\be \label{manus-28082023-23}
\alpha = \frac{1}{(\frac{d-4}{2})!}\,.
\ee
Relations \rf{manus-28082023-20}, \rf{manus-28082023-21}, and \rf{manus-28082023-23} lead to the desired relations given in \rf{manus-19082023-06}.

\appendix{ \large Derivation of gauge transformations of fields \rf{manus-19082023-11} and field strengths \rf{manus-19082023-12}}

In this Appendix, we outline a procedure of the derivation of non-abelian contributions to the gauge transformations of the fields and the fields strengths given \rf{manus-19082023-11} and \rf{manus-19082023-12} respectively. We do this by proving   two statements presented below.

\noinbf{Statement 1}. Let the field strengths $F_{2n+2}^{bc}$ be given by \rf{manus-19082023-05}, while gauge transformations of the vector fields be given by
\be \label{manus-30082023-01}
\delta_\xi \phi_{2n+1}^b = \partial^b \xi_{2n} + \sum_{i=0,\ldots,n} \gamma_{n,i} [\phi_{2n+1-2i}^b,\xi_{2i}]\,,
\ee
where $\gamma_{n,i}$ are unknown coefficients. Let the field strengths $F_{2n+2}^{bc}$ transform in  adjoint representation of the gauge group. Then, firstly, we find the following solution for the coefficients $\gamma_{n,i}$:
\be \label{manus-30082023-02}
\gamma_{n,i} = \alpha_{n,i}\,,
\ee
where $\alpha_{n,i}$ are given in \rf{manus-19082023-06} and, secondly, we find that  gauge transformations of the field strengths $F_{2n+2}^{bc}$ take the form presented in \rf{manus-19082023-12}.

\noinbf{Proof of Statement 1}. Using the gauge transformations of the vector fields \rf{manus-30082023-01}, we find the following gauge transformations of the field strengths $F_{2n+2}^{bc}$ given in \rf{manus-19082023-08}, \rf{manus-19082023-09}
\beq
\label{manus-30082023-03}  && \hspace{-1.2cm} \delta_\xi F_{2n+2}^{bc} =  \partial^b  \sum_{0=1,\ldots,n} \gamma_{n,i}[\phi_{2n+1-2i}^c, \xi_{2i}]
\nonumber\\
&& + \sum_{i=0,\ldots,n} \alpha_{n,i}\Big( [\partial^b\xi_{2i},\phi_{2n+1-2i}^c] + \sum_{j=0,\ldots,i} \gamma_{i,j} [[\phi_{2i+1-2j}^b,\xi_{2i}],\phi_{2n+1-2i}^c]\Big) - (b\leftrightarrow c).\qquad
\eeq
The requirement that the field strengths $F_{2n+2}^{bc}$ transform in adjoint representation of the gauge group implies that the terms proportional to $\partial^b\xi_{2i}$ appearing in \rf{manus-30082023-03} should vanish. It is easy to see that this requirement allows us to fix the coefficients $\gamma_{n,i}$ uniquely as in \rf{manus-30082023-02}. Using \rf{manus-30082023-02} in \rf{manus-30082023-03}, we then get the gauge transformations of the field strengths  $F_{2n+2}^{bc}$ given in \rf{manus-19082023-12}.
In order to cast \rf{manus-30082023-03}  into the form given in \rf{manus-19082023-12} we use the standard Jacobi identity and the relations
\be
\alpha_{n,i+j} \alpha_{i+j,j} = \alpha_{n,j} \alpha_{n-j,i}\,,\hspace{1cm} \alpha_{n,n-i} \alpha_{n-i,j}  = \alpha_{n,j} \alpha_{n-j,i}\,.
\ee

\noinbf{Statement 2}.  Let the field strengths $F^b_{2n+1}$ be given by  \rf{manus-19082023-05}. Let gauge transformations of the vector fields be given by  \rf{manus-30082023-01}, \rf{manus-30082023-02}, while gauge transformations of the scalar fields be given by
\beq
\label{manus-30082023-20} && \delta_\xi \phi_{2n} = - \xi_{2n} + \sum_{i=0,\ldots,n-1} \sigma_{n,i} [\phi_{2n-2i},\xi_{2i}]\,,
\eeq
where $\sigma_{n,i}$ are unknown coefficients. Let the field strengths $F_{2n+1}^b$ transform in adjoint representation of the gauge group. Then, firstly, we find the following solution for $\sigma_{n,i}$:
\be \label{manus-30082023-21}
\sigma_{n,i} = \beta_{n,i}\,,
\ee
where the coefficients $\beta_{n,i}$ are given in \rf{manus-19082023-06} and, secondly, we find that gauge transformations of the field strengths $F_{2n+1}^b$ take the form presented in \rf{manus-19082023-12}.

\noinbf{Proof of Statement 2}. Using the gauge transformation of the vector fields \rf{manus-30082023-01},\rf{manus-30082023-02} and the gauge transformations of the scalar fields \rf{manus-30082023-20}, we find the following gauge transformations of the field strengths $F_{2n+1}^b$ given in \rf{manus-19082023-08}, \rf{manus-19082023-09}:
\beq
\label{manus-30082023-22} && \hspace{-1.2cm} \delta_\xi F_{2n+1}^b =  \sum_{i=0,\ldots,n} \alpha_{n,i}[\phi_{2n+1-2i}^b, \xi_{2i}]  + \partial^b\sum_{i=0,\ldots,n-1} \sigma_{n,i}[\phi_{2n-2i}, \xi_{2i}]
\nonumber\\
&& + \sum_{i=0,\ldots,n-1} \beta_{n,i}\Big( [\partial^b\xi_{2i},\phi_{2n-2i}] + \sum_{j=0,\ldots,i} \alpha_{i,j} [[\phi_{2i+1-2j}^b,\xi_{2i}],\phi_{2n-2i}]\Big)
\nonumber\\
&& + \sum_{i=0,\ldots,n-1} \beta_{n,i}\Big( -[\phi_{2i+1}^b,\xi_{2n-2i}] + \sum_{j=0,\ldots,n-i-1} \sigma_{n-i,j} [\phi_{2i+1}^b,[\phi_{2n-2i-2j},\xi_{2j}]]\Big)\,.\qquad
\eeq
The requirement that the field strengths $F_{2n+1}^b$ transform in adjoint representation of the gauge group implies that the terms proportional to $\partial^b\xi_{2i}$ appearing in \rf{manus-30082023-22} should vanish. It is easy to see that this requirement allows us to fix the coefficients $\sigma_{n,i}$ uniquely as in \rf{manus-30082023-21}. Using \rf{manus-30082023-21} in \rf{manus-30082023-22}, we then get the gauge transformations of the field strengths  $F_{2n+1}^b$ given in \rf{manus-19082023-12}.
In order to cast \rf{manus-30082023-22}  into the form given in \rf{manus-19082023-12} we use the standard Jacobi identity and the relations
\beq
&& \alpha_{n,i} - \beta_{n,n-i} = \beta_{n,i}\,, \hspace{1.8cm}  \alpha_{n,0}=\beta_{n,0}\,,
\\
&& \beta_{n,i+j} \alpha_{i+j,j} = \beta_{n,i} \beta_{n-i,j}\,,\hspace{1cm}
\beta_{n,i} \beta_{n-i,j} = \beta_{n,j} \beta_{n-j,i}\,.
\eeq

\appendix{ \large Derivation of $F^3$-vertex \rf{manus-19082023-50}}

In this Appendix, we outline the derivation of the $F^3$-vertex given in \rf{manus-19082023-50}. To avoid unnecessary technical presentation we outline only main steps of our derivation.

By definition, the $F^3$-vertex is built entirely in terms of the field strengths given in \rf{manus-19082023-05}. The $F^3$-vertex should be invariant with respect to the gauge transformations given in \rf{manus-19082023-12} and the conformal algebra transformations discussed in Sec.\ref{not-conv}. It turns out that the conformal symmetries alone allow us to fix the $F^3$-vertex uniquely. Let us formulate our

\noinbf{Statement}. Requiring that vertex $\LL_{F^3}$ \rf{manus-19082023-50} be invariant under the conformal algebra transformations, we find that the coefficients $c_{n_1,n_2,n_3}^{(1),(2)}$ are fixed uniquely \rf{manus-19082023-51} (up to constant factor).

\noinbf{Proof of the statement}. Firstly, we note that the most general expression for the $F^3$-vertex which is invariant under the Poincar\'e algebra symmetries and the dilatation symmetry is given in \rf{manus-19082023-50}, where the coefficients $c_{n_1,n_2,n_3}^{(1),(2)}$ are arbitrary. Note that the restriction $n_1+n_2+n_3=\frac{d-6}{2}$ in the expression for $\LL^{(1)}$
and the restriction $n_1+n_2+n_3 = \frac{d-4}{2}$ in the expression for $\LL^{(2)}$ are obtained by virtue of the dilatation symmetry. In other words, all that remains is to find restrictions imposed by the conformal boost symmetries.

Secondly, we demonstrate that the conformal boost symmetries allow us to determine the coefficients $c_{n_1,n_2,n_3}^{(1),(2)}$ uniquely.
To this end we note that, by using transformation rules \rf{manus-22082023-33}, it is easy to verify that the requirement of invariance of the $F^3$-vertex under the conformal boost transformations amounts to the requirement of invariance of the $F^3$-vertex under the $R^a$-transformations. In other words, all that we need is to analyse the equation
\be  \label{manus-29082023-01}
\delta_{R^a} \big(\LL^{(1)} - \LL^{(2)})=0\,.
\ee
We now note that the $R^a$-transformations of the field strengths in \rf{manus-19082023-31} can be decomposed as
\be  \label{manus-29082023-02}
\delta_{R^a} = \delta_{R_0^a} + \delta_{R_1^a}\,,
\ee
where $\delta_{R_0^a}$ is the derivative independent part in \rf{manus-19082023-31}, while  $\delta_{R_1^a}$ is the derivative dependent part in \rf{manus-19082023-31}. We then verify that equation \rf{manus-29082023-01} amounts to the following two equations:
\beq
\label{manus-29082023-03} && \delta_{R_1^a} \big(\LL^{(1)} - \LL^{(2)})=0\,,
\\
\label{manus-29082023-04} && \delta_{R_0^a} \big(\LL^{(1)} - \LL^{(2)})=0\,.
\eeq

Thirdly, we consider equation \rf{manus-29082023-03}.
Taking into account that the variation $\delta_{R_1^a} \LL^{(1)}$ depends on three field strengths $F_{2n+2}^{ab}$, while the variation
$\delta_{R_1^a} \LL^{(2)}$ depends on one field strength $F_{2n+2}^{ab}$ and two field strengths $F_{2n+1}^a$, we note that equation \rf{manus-29082023-03} amounts to the following two equations:
\be \label{manus-29082023-05}
\delta_{R_1^a} \LL^{(1)} = 0 \,, \qquad \delta_{R_1^a} \LL^{(2)} =0\,.
\ee
From the first and the second equations in \rf{manus-29082023-05},  we get the following respective two sets of equations for the coefficients $c_{n_1,n_2,n_3}^{(1)}$ and $c_{n_1,n_2,n_3}^{(2)}$:
\beq
\label{manus-29082023-06} && (n_1+1)(d-4-2n_1) c_{n_1+1,n_2,n_3}^{(1)} = (n_2+1)(d-4-2n_2) c_{n_1,n_2+1,n_3}^{(1)}
\nonumber\\
&& \hspace{2cm} = (n_3+1)(d-4-2n_3) c_{n_1,n_2,n_3+1}^{(1)}\,,
\\
\label{manus-29082023-07} && n_1(d-4-2n_1) c_{n_1+1,n_2,n_3}^{(2)} = (n_2+1)(d-4-2n_2) c_{n_1,n_2+1,n_3}^{(2)}
\nonumber\\
&& \hspace{2cm} = n_3(d-4-2n_3) c_{n_1,n_2,n_3+1}^{(2)}\,.
\eeq
The respective solutions to equations \rf{manus-29082023-06} and \rf{manus-29082023-07} are given by
\beq
\label{manus-29082023-08} && c_{n_1,n_2,n_3}^{(1)} = \frac{(\frac{d-4}{2}-n_1)! (\frac{d-4}{2}-n_2)! (\frac{d-4}{2}-n_3)!}{n_1! n_2! n_3!} c_0^{(1)}\,,
\\
\label{manus-29082023-09} && c_{n_1,n_2,n_3}^{(2)} = \frac{(\frac{d-4}{2}-n_1)! (\frac{d-4}{2}-n_2)! (\frac{d-4}{2}-n_3)!}{(n_1-1)! n_2! (n_3-1)!}  c_0^{(2)} \,,
\eeq
where the two coefficients $c_0^{(1)}$, $c_0^{(2)}$ are independent of $n_1$, $n_2$, and $n_3$.

Finally, we consider equation \rf{manus-29082023-04}. Using solutions  \rf{manus-29082023-08}, \rf{manus-29082023-09}, we find that equation \rf{manus-29082023-04} amounts to the relation
\be \label{manus-29082023-10}
c_0^{(1)} = \frac{1}{3}c_0^{(2)}\,.
\ee
Our statement is proved. The normalization for the $c_0^{(1)}$ used in \rf{manus-19082023-51} is a matter of the convention.

\noinbf{Gauge invariance of $F^3$-vertex}. As we have already said, the $F^3$-vertex should be invariant under gauge transformations \rf{manus-19082023-12}. We verified that the $F^3$-vertex with coefficients as in \rf{manus-19082023-51} is indeed gauge invariant. Moreover, the vertices $\LL^{(1)}$ and $\LL^{(2)}$ are separately gauge invariant,
\be
\delta_\xi \LL^{(1)} = 0 \,, \hspace{1cm} \delta_\xi \LL^{(2)} = 0\,.
\ee


\end{document}